%% file: main.tex
\documentclass[sigconf,nonacm,xcolor,dvipsnames,table]{acmart}

\newcommand\vldbdoi{XX.XX/XXX.XX}

\newcommand\vldbpagestyle{plain} 

\usepackage{subfiles}
\usepackage{tikz-cd}
 \usepackage{relsize}
 \usepackage{mathtools}
\usepackage{subcaption}
\usepackage{tabularx}
\usepackage{afterpage}
\usepackage{soul}
\usepackage{balance}
\usepackage{footmisc}
\usepackage[skip=0.5pt]{caption}
\usepackage[export]{adjustbox}
\usepackage[linesnumbered,ruled,noend]{algorithm2e}
\usepackage{listings}
\usepackage{standalone}
\usepackage{enumitem}
\usepackage{amsmath}
\usepackage{pifont}

\newtheorem{observation}{Observation}

\SetCommentSty{mycommfont}

\newcommand{\cmark}{\ding{51}}
\newcommand{\xmark}{\ding{55}}

\newcommand{\optexps}{{\mathcal{E}_c^*}} 
\newcommand{\optexp}{{E_c^*}} 
\newcommand{\optexpsz}{{\mathcal{E}_0^*}}

\newcommand{\eat}[1]{}

\newcommand{\card}[1]{\left|{#1}\right|}

\renewcommand{\epsilon}{\varepsilon}
\renewcommand{\phi}{\varphi}

\newcommand{\sys}{{\textsc{Cluster-Explorer}}}
\newcommand{\system}{\sys}

\AtBeginDocument{%
  \providecommand\BibTeX{{%
    \normalfont B\kern-0.5em{\scshape i\kern-0.25em b}\kern-0.8em\TeX}}}

\newcommand{\reva}[1]{{\leavevmode\color{black}{#1}}}
\newcommand{\revb}[1]{{\leavevmode\color{black}{#1}}}
\newcommand{\revc}[1]{{\leavevmode\color{black}{#1}}}

\newcommand{\revall}[1]{{\leavevmode\color{black}{#1}}}
\newcommand{\fix}[1]{{\leavevmode\color{black}{#1}}}

\begin{document}
\fancyhead{} 

\title{Explaining Black-Box Clustering Pipelines With \system{}}

\setcounter{figure}{0}

\author{Sariel Ofek}
\affiliation{%
  \institution{Bar-Ilan University}
}
\email{sariel.tutay@live.biu.ac.il}

\author{Amit Somech}
\affiliation{%
  \institution{Bar-Ilan University}
}
\email{somecha@cs.biu.ac.il}

\begin{abstract}
Explaining the results of clustering pipelines by unraveling the characteristics of each cluster is a challenging task, often addressed manually through visualizations and queries. Existing solutions from the domain of Explainable Artificial Intelligence (XAI) are largely ineffective for cluster explanations, and interpretable-by-design clustering algorithms may be unsuitable when the clustering algorithm does not fit the data properties.

To bridge this gap, we introduce \system{}, a novel explainability tool for black-box clustering pipelines. Our approach formulates the explanation of clusters as the identification of concise conjunctions of predicates that maximize the coverage of the cluster's data points while minimizing separation from other clusters. We achieve this by reducing the problem to generalized frequent-itemsets mining (gFIM), where items correspond to explanation predicates, and itemset frequency indicates coverage. To enhance efficiency, we leverage inherent problem properties and implement attribute selection to further reduce computational costs. Experimental evaluations on a benchmark collection of 98 clustering results demonstrate the superiority of \system{} in both explanation quality and execution times compared to XAI baselines.

\end{abstract}

\maketitle
\setcounter{page}{1}

\pagestyle{\vldbpagestyle}




\input{introduction}

\input{related}

\input{solution }

\input{algorithm}

\input{experiments}

\input{conclusions}

\clearpage

\bibliographystyle{ACM-Reference-Format}
 \small{
 \bibliography{bibtex}
 }  
\end{document}

%% file: introduction.tex
\section{introduction}

Cluster analysis is an important data mining tool widely used to segment data points into meaningful groups (clusters) in an unsupervised manner, without the need for labeled data.

Similar to the development of a machine learning (ML) predictive model, data scientists optimize clustering outcomes by employing \textit{clustering pipelines}—a series of data preprocessing and preparation steps (e.g., scaling, transformations, dimensionality reduction) followed by the application of a clustering algorithm such as K-means, spectral clustering, or affinity propagation. The choice of algorithm often depends on the data properties and application domain~\cite{xu2015comprehensive,ezugwu2022comprehensive}.

\begin{figure}[t]
    \vspace{+7mm}
    \includegraphics[width=0.8\columnwidth]{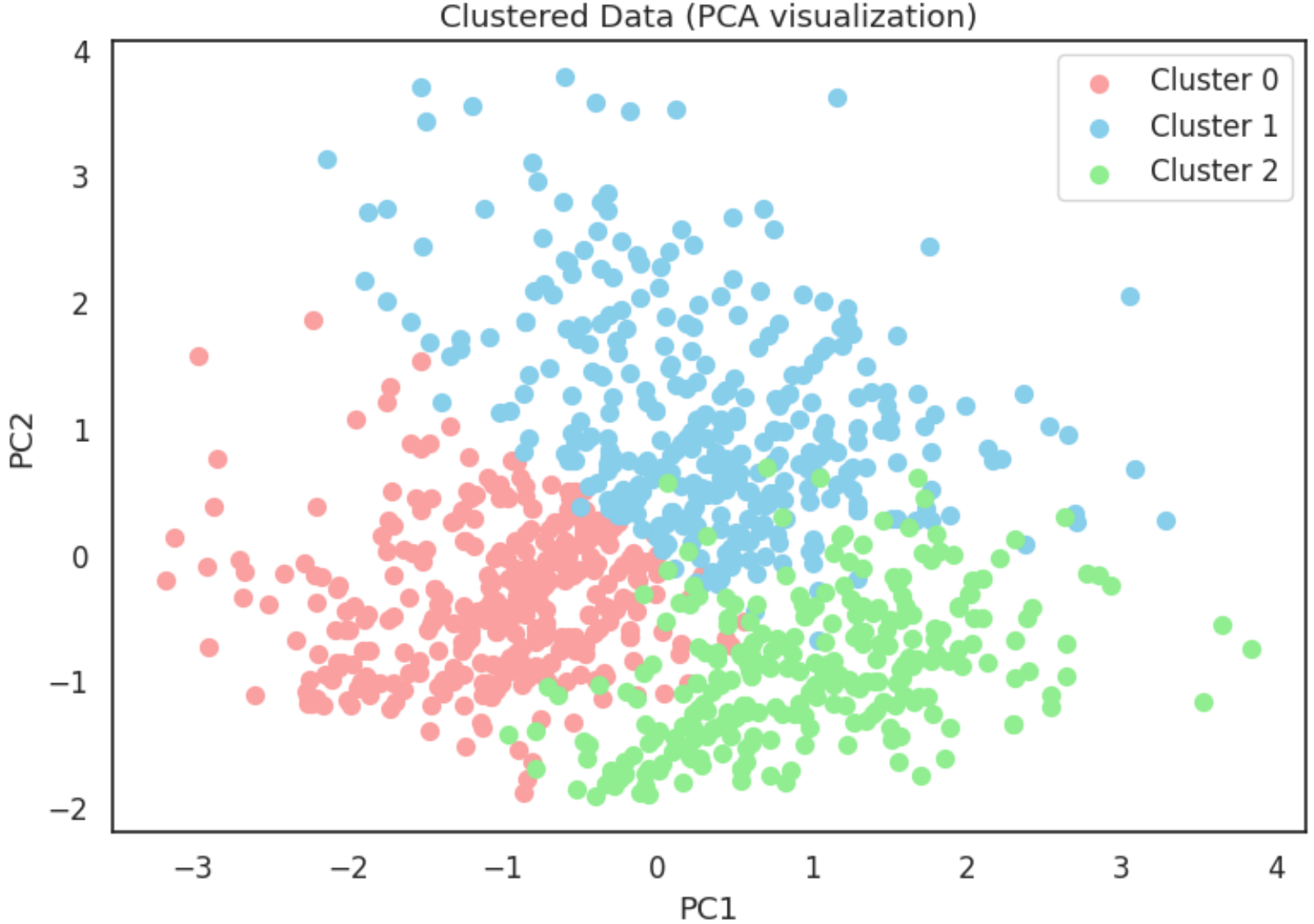}
    \vspace{+1mm}    
    \caption{Clustering results visualization (Adult dataset)}
    \vspace{+1mm}
    \label{fig:clusters_vis}
\end{figure}

While the results of these algorithms can be easily visualized on a two-dimensional plane (see Figure~\ref{fig:clusters_vis}), allowing users to inspect how well the data points are separated, interpreting the \textit{meaning} of the segmentation and understanding the characteristics of each cluster is challenging. This process often requires users to manually perform additional analytical queries and subsequent data visualizations on the clustered data. For illustration, consider the following example.

\begin{table}[t!]
\centering
\resizebox{\columnwidth}{!}{
\small
\ttfamily
    \begin{tabular}{|c|c|c|l|l|c|c|c|c|}
    \hline
    \rowcolor{gray!40}
    \textbf{Row ID} & \textbf{Age} & \textbf{Edu.num} & \textbf{Relationship} & \textbf{Gender} & \ldots &\textbf{Hrs-per-week} & \textbf{Income} &  \textbf{Cluster}\\
    \hline
    
    124 & 25 & 7 & Unmarried  & Male & \ldots & 40 & $\leq 50K$ & 0 \\
    32 & 41 & 10 & Unmarried  & Female & \ldots & 50 & $\geq 50K$ & 0 \\
    53 & 34 & 12 & Unmarried  & Male & \ldots & 50 & $\geq 50K$ & 0 \\
    \ldots & \ldots &\ldots &\ldots &\ldots &\ldots &\ldots &\ldots &\ldots  \\
    342 & 36 & 3 & Husband & Male & \ldots & 50 & $\geq 50K$ & 1 \\
    521 & 40 & 3 & Husband & Male & \ldots & 50 & $\leq 50K$ & 1 \\
    5631 & 45 & 5 & Wife & Female & \ldots & 60 & $\geq 50K$ & 1 \\
    \ldots & \ldots &\ldots &\ldots &\ldots &\ldots &\ldots &\ldots &\ldots  \\
    39 & 46 & 12 & Wife & Female & \ldots & 30 & $\leq 50K$ & 2 \\
    938 & 33 & 15 & Husband & Male & \ldots & 60 & $\geq 50K$ & 2 \\
    693 & 36 & 14 & Husband & Male & \ldots & 50 & $\geq 50K$ & 2 \\
    \ldots & \ldots &\ldots &\ldots &\ldots &\ldots &\ldots &\ldots &\ldots  \\
    \hline
\end{tabular}}
\caption{Adult dataset sample with cluster labels}
\vspace{-3mm}
\label{tab:adults}    
\end{table}

\begin{figure*}[t]
\vspace{-6mm}
    \includegraphics[width=1\textwidth]
    {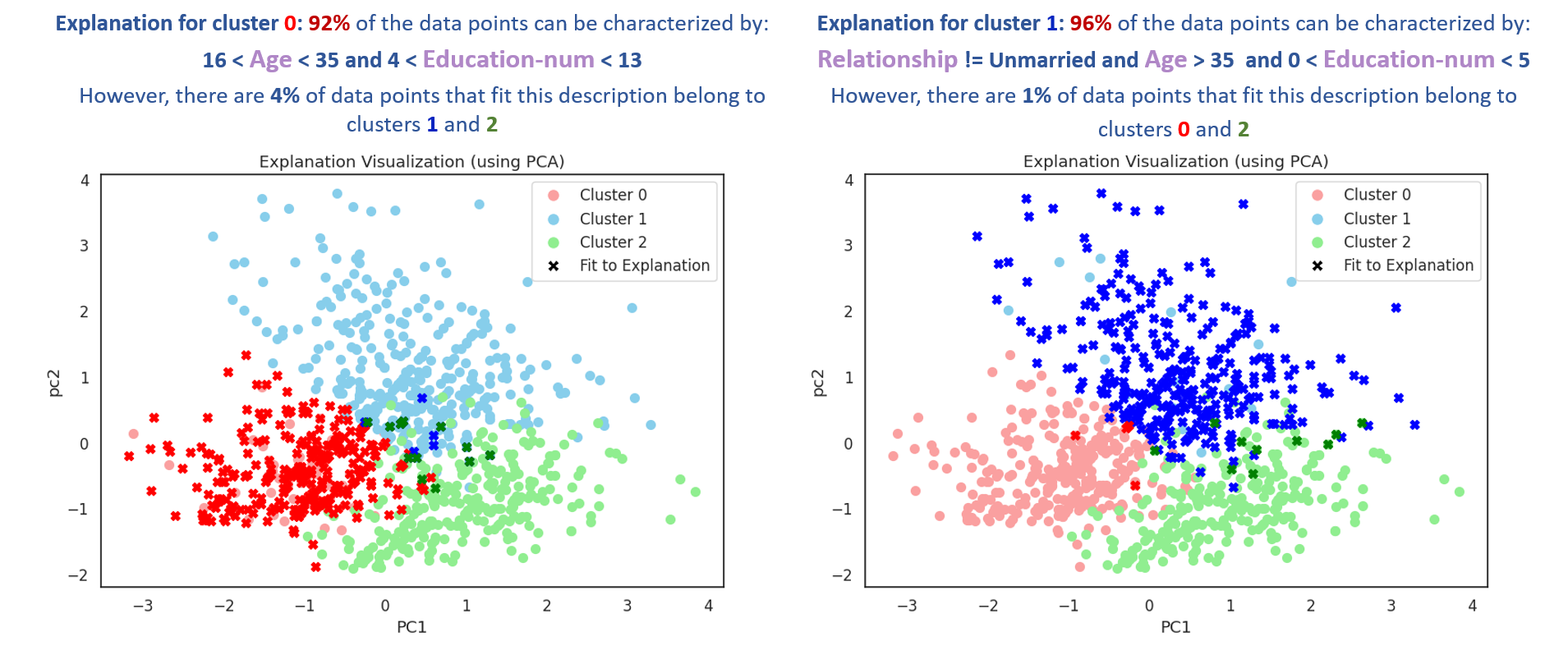}
    \vspace{-3mm}
    \caption{Example cluster explanations generated by \system{} }
    \label{fig:screenshot}
\vspace{-2mm}
\end{figure*}

\begin{example}\label{ex:bad_example}
Consider Clarice, a data analyst examining the well-known ``Adult'' income dataset\cite{adults_dataset}, which contains demographic information on individuals alongside their income (see Table~\ref{tab:adults} for a sample). Clarice employs a clustering pipeline that includes one-hot encoding of categorical data, Z-scaling of numerical columns, and dimensionality reduction using PCA~\cite{dunteman1989principal}. She then applies an agglomerative (hierarchical) clustering algorithm~\cite{mullner2011modern} to the processed data and visually examines the results, as illustrated in Figure~\ref{fig:clusters_vis}. She sees that the data points are fairly segmented into three clusters. However, it is still unclear clear what the characteristics of each cluster are. Specifically, what attributes are shared among points within the same cluster, and what properties differentiate the clusters from one another?
\end{example}

Previous work~\cite{shrikumar2017learning,sundararajan2017axiomatic,ribeiro2016should,lundberg2017unified,ribeiro2018anchors,linardatos2021explainable} in the domain of Explainable Artificial Intelligence (XAI) has primarily focused on \revc{post-hoc explanations of the outcome of supervised models}. This is often achieved by calculating importance scores for features~\cite{zien2009feature,ibrahim2019global} or feature-value combinations~\cite{saarela2021comparison,ribeiro2016should,lundberg2017unified}. To apply these methods for interpreting cluster results, one must fit an auxiliary supervised model on the clustering labels and aggregate the explanations for each cluster, which is a nontrivial task. \revb{Another line of research focuses on interactive, visual tools for cluster analysis~\cite{kandogan2012just,cavallo2018clustrophile,kwon2017clustervision}, enabling the construction and basic evaluation of clustering pipelines without the need for coding or SQL expertise.
}
Closer to our work, systems such as~\cite{frost2020exkmc,moshkovitz2020explainable,kauffmann2022clustering,gad2020excut} recognizes the importance of explaining clusters and proposes clustering algorithms that are \textit{interpretable by design}. However, it is widely accepted~\cite{xu2015comprehensive,ezugwu2022comprehensive} that each clustering algorithm is suitable for a specific data domain and properties. Therefore, there is a significant need for a framework that can produce explanations for any given clustering pipeline.

To this end, we present \system{}, a system for \revc{post-hoc explanations for black-box clustering pipelines}. Given the original dataset and the clustering pipeline results, \system{} automatically generates coherent explanations that characterize each cluster. We define a cluster explanation as a conjunction of predicates (e.g., $\langle \text{`Age'}, \text{between}, \text{(16,35)}\rangle \wedge \langle \text{`relationship'}, \neq, \text{Husband}\rangle$), and use an efficient algorithm to generate a set of explanations for each cluster that optimizes the following criteria, based on XAI Explanation principles~\cite{miller2019explanation,wang2019designing}: (1) high \textit{cluster coverage} — the explanation should describe as many of the cluster's data points as possible; (2) low \textit{separation error} — the explanation should apply to a minimal number of data points from other clusters; and (3) high \textit{conciseness} — the explanation should be brief, comprising a minimal number of predicates, to ensure coherency and applicability~\cite{miller2019explanation}.

\begin{example}\label{ex:good_example}
Figure~\ref{fig:clusters_vis} depicts two example cluster explanations generated by \system{} for the clustering pipeline results described in Example~\ref{ex:bad_example}. Cluster 0 (left hand side) is characterized by individuals with 'Age' between 16 and 35 and 'Education-num' between 4 and 13 (i.e., from middle school education level up to a bachelor's degree). Note that this explanation is not ``perfect''—it covers 92\% of the cluster points, but 4\% of the data points it covers belong to different clusters (as indicated by the X marks on the left-hand side of Figure~\ref{fig:clusters_vis}).

The explanation for Cluster 1 is different, primarily characterized by older individuals over 35, with a middle school (or lower) education level, and a relationship status that is not `unmarried'. This explanations is slightly longer, but covers 96\% of the cluster's data points with only a 1\% error. 
\end{example}

To efficiently generate such explanations, we use a reduction to the problem of generalized frequent itemsets mining~\cite{gfim_orig,gfim1,gfim2,gfim3} (gFIM). A gFIM algorithm operates on a \textit{transactional} dataset, where each transaction comprises a set of discrete items, and the items are associated with categories in an additionally provided taxonomy. The result is a set of \textit{generalized} frequent itemsets, containing either items or categories that include them. Intuitively, in our problem, the items are equivalent to explanation predicates, and the frequency of itemsets corresponds to the explanations' \textit{coverage}.

Before employing the gFIM algorithm, we transform the raw data into a set of \textit{augmented transactions}, with the goal of increasing and enriching the set of predicates that can be used in the cluster explanations, beyond the raw values. Specifically, we use multiple binning methods (e.g., equal width, 1-D clustering, tree-based binning, etc.) for each numeric attribute and further augment categorical values with negation predicates for the rest of the values not appearing in the row.
Once the data is transformed, we organize all numerical bins in a taxonomy of intervals. We then execute the gFIM algorithm separately for each cluster, using the taxonomy to eliminate the possibility of overlapping predicates and efficiently find minimal-size sets of predicates that maximize the coverage of each cluster's data points.
To obtain the final set of explanations for each cluster, we further process the gFIM output by filtering out explanations with high separation error, and calculating a set of Pareto Optimal~\cite{censor1977pareto,borzsony2001skyline} explanations, where each explanation demonstrates an optimal trade-off between the three criteria for explanation quality.

However, gFIM algorithms are known for their lack of scalability, as their cost can be exponential w.r.t. the number of items. In \system{}, we tackle this issue in two ways:
First, since we aim for explanations with high coverage (accounting for the majority of a cluster's data points) and small size, we leverage these natural properties to restrict the execution of the gFIM algorithm. By confining the gFIM algorithm to highly frequent itemsets of small size, we significantly accelerate running times, as most infrequent itemsets are pruned in the early stages of execution.

Second, we introduce a simple yet highly effective attribute selection optimization based on feature importance calculation~\cite{zien2009feature} using a set of decision tree models fitted separately for each cluster. By selectively limiting the computation to promising attributes, the gFIM algorithm operates on significantly fewer items, a crucial factor affecting its performance~\cite{yang2004complexity}.

An extensive set of experiments was conducted to evaluate \system{}. We first devised a benchmark dataset containing 98 clustering instances, resulting from the execution of 16 different clustering pipelines with 5 different algorithms on 19 source datasets. We further examined the quality of the clusters by calculating the \textit{silhouette coefficient}~\cite{rousseeuw1987silhouettes}, filtering out poor clustering pipeline results. We compared the quality of the explanations according to coverage, separation error, and conciseness, as well as the running time of \system{}, against four different baseline approaches from the domain of XAI.

Our results show that the explanations generated by \system{} are superior to those of the baselines in terms of both quality and running times. Additionally, we demonstrate that our attribute selection optimization improves running times by an average of 14.4X, with a negligible decrease in explanation quality.

A prototype of our solution, wrapped with a user interface, was recently demonstrated in~\cite{tutay2023cluster}. The accompanying short paper briefly presents the problem and outlines our solution, but it lacks significant algorithmic and optimization details, as well as an experimental analysis which are provided in this paper.

\vspace{1mm}
Our main contributions in this paper include:
\begin{itemize}
\item We introduce \system{}, a framework for \revc{post-hoc explanations of black-box clustering pipelines}.
\item We develop an efficient algorithm based on a careful reduction to generalized frequent itemsets mining~\cite{gfim_orig}, combined with a predicate-augmentation process and a dedicated attribute selection method, enabling \system{} to efficiently mine concise explanations with good coverage of each cluster's data points.
\item We create a benchmark dataset of 98 clustering results, curated from 16 clustering pipelines using 5 different algorithms and 19 datasets (publicly available in~\cite{our_github}).
\item We implement a prototype of \system{}, also publicly available in~\cite{our_github}, and conduct extensive experiments demonstrating the superiority of our approach in explanation quality and running time.
\end{itemize}

\paragraph*{Paper Outline}
We survey related work in Section~\ref{sec:related},  then describe our model and problem definition in Section~\ref{section:model}. We present our algorithms in Section~\ref{section:algorithm}, experiments in~\ref{sec:experiments}, and close in Section~\ref{sec:conclusion}.



%% file: related.tex
\section{Related Work}
\label{sec:related} 

We next survey relevant lines of related work.

\paragraph*{Cluster Analysis and Algorithms}
A plethora of unsupervised clustering algorithms have been proposed~\cite{kmeans_paper,dbscan_paper,zhang1996birch,spectral_paper,affinity_propagation}, each often suitable for different data properties and application domains~\cite{xu2015comprehensive,ezugwu2022comprehensive}. As with many data mining and machine learning processes, data preparation and preprocessing steps, such as imputing missing values, scaling, and reducing data dimensionality, are crucial for optimizing the results of clustering algorithms~\cite{garcia2016big,alasadi2017review,zhang2003data}.  

\fix{However, as noted, characterizing and understanding the resulted clusters is a challenge, often requires the user to employ subsequent analytical operations to explore the differences between the data points in each resulting cluster.}
\system{} is specifically designed to address this challenge by explaining the results of black-box clustering pipelines, which may use any clustering algorithm, using a set of coherent and concise rule-like explanations generated for each cluster.



\revb{\paragraph*{Interactive Visual Interfaces for Cluster Analysis}
An adjacent line of work focuses on developing visual, interactive tools for cluster analysis, contributing to broader efforts in designing data exploration interfaces that eliminate the need for coding skills or SQL expertise~\cite{yuan2021survey}. For instance, tools like~\cite{cavallo2018clustrophile,kwon2017clustervision,kandogan2012just} allow users to interactively refine the clustering pipeline by selecting different algorithms and dimensionality reduction techniques, while providing basic descriptive statistics and annotations for the resulting clusters. Other works, such as~\cite{xia2022interactive,chatzimparmpas2020t}, specifically help users investigate the outcomes of dimensionality reduction methods, offering visual tools to inspect inaccuracies and understand the similarities between points in the low-dimensional projection space.

Unlike these approaches, \system{} tackles a complementary task: explaining each resulting cluster by discovering a concise set of rules that tightly characterizes the cluster. \system{} can be used side-by-side with visual cluster analysis interfaces, enriching the interactive process by offering robust explanations to help users refine their cluster analysis.
}

\paragraph*{XAI, Explainable ML}
\revc{
Numerous previous works propose solutions for explaining the predictions of supervised ML models. While some approaches focus on developing explainable-by-design ML models~\cite{arrieta2020explainable} (see more below), a prominent line of research emphasizes \textit{post-hoc} analysis of model predictions, where the explainer does not require access to the internal workings of the model ~\cite{shrikumar2017learning,sundararajan2017axiomatic,ribeiro2016should,lundberg2017unified,ribeiro2018anchors} (see \cite{linardatos2021explainable} for a survey).
}. A key approach in this subfield explains individual predictions by assessing \textit{importance}~\cite{saarela2021comparison,ribeiro2016should,lundberg2017unified} for each feature-value pair of a data point $x$ regarding the prediction $M(x)$. Such explanations are called \textit{local}, while \textit{global} explanations~\cite{zien2009feature,ibrahim2019global} measure feature importance for the model's overall behavior. Closer to our work \cite{ellis2021algorithm} introduces a feature importance tool for clustering, akin to~\cite{lundberg2017unified}.
Another line of research focuses on extracting \textit{decision rules} from complex ML models. Works like~\cite{ribeiro2018anchors,lore_guidotti2019factual} generate if-then rules for individual predictions, highlighting how specific attribute changes influence outcomes, whereas global explanation approaches~\cite{friedman2008predictive,cohen1999simple} mine rules from ensembles~\cite{sagi2018ensemble,freund1997decision} to enhance model transparency. 
A recent study~\cite{bobek2022enhancing} applies Anchor~\cite{ribeiro2018anchors} explanations to cluster samples. 

However, as demonstrated in our experiments, aggregating \textit{local} explainers~\cite{lundberg2017unified,ribeiro2018anchors} and extracting rule-based cluster explanations from ensemble models trained on cluster labels~\cite{sagi2018ensemble,freund1997decision} yield suboptimal results due to overfitting~\cite{hawkins2004problem,ying2019overview} and high computational costs. In contrast, \system{} does not rely on supervised ML models for explanations and efficiently generates high-quality, rule-like explanations.

\paragraph*{Interpretable-by-design Clustering Algorithms}
Numerous previous works recognize the difficulty in interpreting clustering results~\cite{frost2020exkmc,moshkovitz2020explainable,kauffmann2022clustering,gad2020excut} and address this issue by suggesting clustering algorithms that are interpretable by design. For example,~\cite{frost2020exkmc,moshkovitz2020explainable} propose approximating the K-means algorithm using decision trees; \cite{kauffmann2022clustering} introduces a neural network formulation of K-means, allowing the use of explainability techniques for networks (e.g., class activation maps~\cite{zhou2016learning}); and \cite{gad2020excut} presents a specialized solution for clustering knowledge graph entities, which performs an embedding-based clustering process combined with rule-based explanation mining to cover the majority of entities in each cluster.

However, all of these methods are tailored to a specific, single clustering algorithm, which, as mentioned above, may be ineffective for some datasets and applications~\cite{xu2015comprehensive,ezugwu2022comprehensive}. In contrast, \system{} can produce explanations for any given clustering pipeline, supporting a variety of clustering algorithms.

%% file: solution.tex

\section{Model \& Problem Definition}
\label{section:model}

\subsection{Data model, Explanation Candidates}
\label{ssec:model}

We next describe a data model for clustering pipelines then define candidate explanations for clustering results.

\paragraph*{Clustering Pipeline} We assume an input dataset $D = \langle X, A \rangle$, containing $n$ data points $X = x_1, \dots, x_n$,
projected over $m$ attributes $A = a_1, \dots, a_m$.
We denote by $x_i$ the $i$-th data point, and by $x_{i,a}$ the projection of $x_i$ over attribute $a$.
A clustering pipeline is then defined as a series of data transformations applied to $D$ (e.g., normalization, null-value imputation, one-hot encoding, dimensionality reduction, etc.). These transformations result in a modified dataset $D' = \langle X, A' \rangle$, where the data points $x_1, \dots, x_n$ are projected onto a transformed attribute space $A' = a_1, \dots, a_{m'}$. A clustering algorithm is then applied to $D'$, resulting in a clustering mapping function $CL: X \rightarrow C$, which associates each data point $x_i$ with a cluster $c \in C$, where $C$ is a set of cluster labels.


\paragraph{Cluster Explanation Candidates.} Following XAI works for producing explanations for ML models~\cite{ribeiro2018anchors, lore_guidotti2019factual}, which highlight the usefulness of rules-like explanations, we define an explanation, denoted $E$, as a conjunction of predicates $E=\{P_1 \wedge P_2\dots P_l \}$. The predicates are of the form $P \coloneqq \langle a, op, V \rangle$, where $a\in A$, $op$ is an operator (e.g., <, >, `contains', `between', 'not', ...), and $V$ is a set of literal values.
Given a data point $x \in X$, we say that an explanation $E$ \textit{holds} for $x$ if $x$ satisfies all the predicates in $E$. Specifically, $E(X) = \text{true} \iff \forall P \in E, P(x) = \text{true}$.

When considering explanations in the context of clustering results, naturally, a subpar explanation for cluster $c$, denoted $E_c$, may hold for data points labeled as $c$ as well as for points assigned to different clusters $c' \in C$ with $c' \neq c$. In Section~\ref{ssec:measures}, we describe three criteria for selecting effective cluster explanations.

\begin{table*}[t]
\vspace{-3mm}
\resizebox{\textwidth}{!}{
\centering
\ttfamily
{ %
    \begin{tabular}{|c|c|c|c|l|l|}
    \hline
    \textbf{Exp.num} & \textbf{Explanation Candidate} & \textbf{Cluster label} & \textbf{\textit{coverage}} & \textbf{\textit{Separation Error}} & \textbf{Conciseness} \\
    \hline
    $E_0^1$ &  \begin{tabular}{@{}c@{}}   $\langle \text{‘age’,between,(16,48)} \rangle \wedge \langle \text{'education-num’,between,(4,13)} \rangle $\\  $\wedge \langle \text{'relationship’,!=,Husband} \rangle $\end{tabular}
    
     &  \text{0} & \text{0.99} &  0.05 &  0.33  \\
    \hline
    $E_0^2$ & \begin{tabular}{@{}c@{}}  $\langle \text{‘age’,between,(16,35)} \rangle \wedge $ \\  $\langle \text{'education-num’,between,(4,13)} \rangle $\end{tabular} &  \text{0} & 0.95 &  0.04 &  0.5 \\
    \hline
    $E_0^3$ &  \begin{tabular}{@{}c@{}}  $  \langle \text{‘age’,between,(16,53)}\rangle \wedge \langle \text{'hours-per-week’,between,(10,72)} \rangle$ \\ $ \wedge  \langle\text{'education-num’,between,(4,14)} \rangle$\end{tabular} &   \text{0} & 0.88 &  0.04 &  0.33 \\
    \hline
\end{tabular}}
}%
\captionof{table}{Example Candidate Explanations}
\label{tab:explanations}
\vspace{-2mm}
\end{table*}

\begin{example}
Consider again Table~\ref{tab:adults}, with a sample of the raw Adult dataset, alongside resulted cluster labels, as described in Example ~\ref{ex:bad_example}. 
See that, for example, the first three rows in the table (IDs 124, 32, and 53) are labeled as Cluster 0. 
Now, three candidate explanations for Cluster 0 are depicted in Table~\ref{tab:explanations} (ignore, for now, the three right-most columns).
Explanation $E_0^2$, for example, comprises of two predicates:

$P_1 \coloneqq \langle \text{`age'}, between, (16, 35) \rangle$, and 

$P_2 \coloneqq \langle \text{`education-num'}, between, (4, 13)\rangle$.

Out of the rows in Table~\ref{tab:adults} that indeed belong to Cluster 0, see that Explanation $E_0^2$ holds for Rows 124 and 53,
yet is not true for Row 32 (having \textit{age} value of 41). 
By contrast, Explanation $E_0^1$ holds for all three rows (IDs 124, 32, and 53) yet unfortunately -- it also holds for rows 5631 and 39, which are assigned to different clusters (Cluster 1 and Cluster 2).
\end{example}

To generate effective explanations, we next define three measures for evaluating the coverage, separation, and conciseness of cluster explanation candidates.

\subsection{Quality Measures for Cluster Explanations}
\label{ssec:measures}
The question of what constitutes a ‘good’ explanation has been investigated in various different domains such as cognitive science, philosophy and psychology. Relying on this vast body of research, works e.g. \cite{miller2019explanation,wang2019designing} suggest that in the context of XAI, a \textit{good} explanation is primarily \textit{contrastive} (i.e., why event $P$ happened \textit{instead} of an event $Q$), but also \textit{simple}, \textit{coherent}, and \textit{truthful}. 

In \system{}, we adapt these criteria to the use case of explaining clustering results and develop corresponding quality metrics for explanations. Given a cluster $c \in C$, an ideal explanation $E_c$ should have (1) high \textit{coverage} of the points in $c$, while maintaining a (2) low \textit{separation} error. Specifically, the explanation must be valid for the majority of data points in cluster $c$, and invalid for data points associated with any other cluster $c' \in C$, where $c' \neq c$. The higher the scores with respect to (1) and (2), the more contrastive and truthful the explanation is. 
Similar measures to (1) and (2) have been previously proposed in the context of evaluating decision rules~\cite{ribeiro2018anchors,lore_guidotti2019factual}, \revb{drawing parallels to the notions of precision and recall commonly used in supervised learning.}
 
However, as noted previously, a naive explanation achieving perfect scores might consist of the union of all attribute-value pairs for each data point in $c$. Naturally, such naive explanations are be overly verbose and incoherent. To address this, we introduce a (3) \textit{conciseness} measure, which considers the number of predicates in the explanation $E_c$. We next provide formal definitions for these three quality measures of explanations.

\noindent\textbf{1. Cluster Coverage.}
Given a set of data points $X$, cluster labels $C$, and a clustering mapping function $CL$, the \textit{coverage} of an explanation $E_c$ (for cluster $c$) is defined as the ratio of the points in cluster $c$ for which explanation $E_c$ holds:
$$
Coverage({E_c}) \coloneqq 
\frac{ \card{\{x\in X\mid E_c(x) = true \wedge CL(x)=c\}}}{\card{\{x\in X\mid CL(x)=c\} }}
$$

 \noindent\textbf{2. Separation Error.}
This measure is defined as the ratio of points for which the explanation $E_c$ holds, yet these points do not belong to cluster $c$:
\begin{equation*}
\setlength{\abovedisplayskip}{2mm}
\setlength{\belowdisplayskip}{2mm}
SeparationErr(E_c) := \frac{\card{\left\{x\in X\mid E_c(x) = True \wedge CL(x)\in C \setminus \{c\} \right\}}}{\card{\{x\in X\mid E(x) = true}}
\end{equation*}

\noindent\textbf{3. Conciseness.} Following~\cite{miller2019explanation}, the length and simplicity of the explanations are important for its comprehension by the users. We therefore define the conciseness of an explanation to be the inverse number of predicates it contains:
\begin{equation*}
 \setlength{\abovedisplayskip}{2mm}
\setlength{\belowdisplayskip}{2mm}
Conciseness(E_c) := \frac{1}{\card{\{P~|~P~\text{is a predicate in $E_c$} \} }}
\end{equation*}

\vspace{1mm} The following example demonstrates the three measures measures for the explanations in Table~\ref{tab:explanations}. 
\begin{example}
    Consider Explanation $E_0^1$ for Cluster 0, as in Table~\ref{tab:explanations}. Assume that the number of data points belonging to Cluster 0 is 373, out of which 370 satisfy $E_0^1$. In addition, 20 other data points that belong to Clusters 1 and 2 also satisfy $E_0^1$. 
    Calculating the scores for $E_0^1$ we obtain:
    $Coverage({E_0^1)}) = \frac{370}{373} = 0.99$,  $SeparationErr({E_0^1)}) = \frac{20}{390} = 0.05$, and $Conciseness({E_0^1)}) = \frac{1}{3} = 0.33$.   
    
\end{example}

\subsection{Problem Definition}
There is a natural trade-off between the quality measures. For instance, an explanation obtaining a very high \textit{coverage}, may have a lower \textit{conciseness} score and higher \textit{separation error}, whereas a highly \textit{concise} explanation may fall short on \textit{coverage}. 

We therefore define the problem of generating cluster explanations as follows: 
Given user-defined thresholds for coverage, separation error, and conciseness, we aim to find a set of desired explanations for a cluster $c$, 
denoted by $\optexps$, such that each $\optexp \in \optexps$ (1) meets the 
thresholds criteria, and (2) is \textit{Pareto optimal}~\cite{censor1977pareto}, meaning that there is no other explanation that surpasses $\optexp$ with respect to all three measures. Formally, the problem is defined as follows.

\begin{definition}[Cluster Explanations Generation Problem]
For data points $X$, a cluster $c \in C$, we define the set of desired explanations $\optexps$ using the following two criteria: 
\begin{equation}
\setlength{\abovedisplayskip}{2mm}
\setlength{\belowdisplayskip}{2mm}
\begin{aligned}
\forall \optexp \in \optexps,~~ 
& Coverage(\optexp) \geq \theta_{cov}  \\ 
& \wedge~ SeparationErr(\optexp) \leq \theta_{sep} \\  
& \wedge~ Conciseness(\optexp) \geq \theta_{con}  
\end{aligned}
\end{equation}
\begin{equation}
\setlength{\abovedisplayskip}{2mm}
\setlength{\belowdisplayskip}{2mm}
\begin{aligned}
\forall \optexp \in \optexps~~\nexists E_c,~~ & Coverage(E_c) \geq Coverage(\optexp) \\
 & \wedge  SeparationErr(E_c) \leq SeparationErr(\optexp) \\  & \wedge Conciseness(E_c) \geq Conciseness(\optexp)   
\end{aligned} 
\end{equation}

\end{definition}

\begin{example}
    Consider again the candidate explanations depicted in Table~\ref{tab:explanations}. We aim to generate $\optexpsz$, i.e., the set of optimal explanations for Cluster 0, with the coverage, separation, and conciseness thresholds defined as $\theta_{cov} = 0.8$, $\theta_{sep} = 0.05$, and $\theta_{con} = 0.33$. First, note that all explanations $E_0^1$, $E_0^2$, and $E_0^3$ meet the threshold criteria. Regarding Pareto optimality, Explanation $E_0^1$ surpasses $E_0^2$ with respect to coverage (0.99 compared to 0.95) but is inferior with respect to separation error (0.05 compared to 0.04) and conciseness (0.33 compared to 0.5). As for $E_0^3$, it is \textit{dominated} by $E_0^2$, having the same separation error (0.04) yet better coverage and conciseness (0.95 compared to 0.88, and 0.5 compared to 0.33). Therefore, $\optexpsz = \{E_0^1, E_0^2\}$.

\end{example}

We next devise an efficient algorithm, based on a reduction of our problem to generalized frequent itemset mining (gFIM) in order to generate the explanation set $\optexps$ for each cluster $c \in C$.

%% file: algorithm.tex
\section{Algorithm}
\label{section:algorithm}

We next describe the cluster explanation generation process performed by \system{}. The effectiveness and efficiency of our solution stem from a careful reduction to the problem of generalized frequent itemset mining~\cite{gfim_orig,gfim1,gfim2,gfim3} (gFIM). In our context, gFIM is used to mine sets of predicates that concisely characterize the majority of a cluster's data points. We first provide a brief background on the gFIM problem and outline our approach, followed by a detailed discussion of each phase.

\subsection{Background \& Algorithm Outline}

\paragraph*{Generalized Frequent Itemsets Mining (gFIM)}
This problem extends the classic data mining problem of finding frequent itemsets (and association rules) in transactional data~\cite{agrawal1994fast,han2000mining}. 
Given a set of items $I = \{A, b, 1, 2\}$, let $T =\{[A,1], [b,2], [A,2] \}$ be a set of transactions. Given a \textit{support} (frequency) threshold of $2/3$, we see that the itemsets $\{A\}$ and $\{2\}$ are frequent (i.e., occur in two out of three transactions).  
In the extended problem of \textit{generalized} frequent itemsets mining, a taxonomy of of item's \textit{categories} is also provided as input, then the mined generalized frequent itemsets may contain either an item or one of its associated categories.  
For example, given the following taxonomy for items $I$: 
\vspace{-1mm}
\[
\begin{tikzcd}[column sep=small, row sep=small]
\textit{Character} \arrow[r] \arrow[rd] & \textit{  Number} & ~~~\textit{       Lowercase} \\
 & \textit{Letter} \arrow[r] \arrow[ur] & \textit{Capital} \\
\end{tikzcd}
\vspace{-4mm}
\]

We can now obtain generalized frequent itemsets from $T$ such as  $\{\textit{Letter},2\}$ and $\{A,\textit{Number}\}$, both has a frequency of $2/3$.
Multiple algorithms~\cite{gfim_orig,gfim1,gfim2,gfim3} can be used for efficiently mining such itemsets, given a support threshold and maximal desired itemset size. Naturally, the higher the input support threshold and the lower is the maximal itemset size -- the better are the performance.

\paragraph*{Cluster Explanations Algorithm Overview}
Intuitively, in our problem, the items correspond to explanation predicates, and the frequency of itemsets is equivalent to the explanations' \textit{coverage}. As we explain in Section~\ref{ssec:transactional}, before employing the gFIM algorithm, we first transform the raw data into a set of augmented transactions. Each numeric attribute in a data point is binned using multiple binning strategies, and categorical values are augmented with negation predicates for values not appearing in the row. Once the data is transformed, we organize all numerical bins into an interval taxonomy based on their containment partial order. The goal of our augmentation is to increase and enrich the set of predicates that can be used in the cluster explanations, extending beyond the original value domain.

As detailed in Section~\ref{ssec:gfim}, we then apply the gFIM algorithm to the augmented transactions and interval taxonomy for each cluster. This allows us to efficiently dice the predicates hierarchy and exclude overlapping or suboptimal explanation predicates. Specifically, the mining process is accelerated by using a high support threshold, $\theta_{cov}$, which indicates that explanations should cover the vast majority of the data points, and a relatively small maximal itemset size, as effective explanations should be \textit{concise}~\cite{miller2019explanation}. The candidate explanations generated by the gFIM algorithm are further filtered to remove those with high separation error. We then select only the Pareto optimal explanations using the skyline operator~\cite{borzsony2001skyline}.

Finally, in Section~\ref{ssec:sampling}, we describe a simple yet highly effective optimization technique that reduces running times by limiting the number of attributes, a known factor that significantly affects the cost of gFIM when applied on relational data~\cite{fu1995meta}.

\subsection{Transforming the raw data to a set of augmented transactions}
\label{ssec:transactional}

\begin{figure}[t]
\vspace{-4mm}
\centering
\includegraphics[width=0.98\columnwidth]{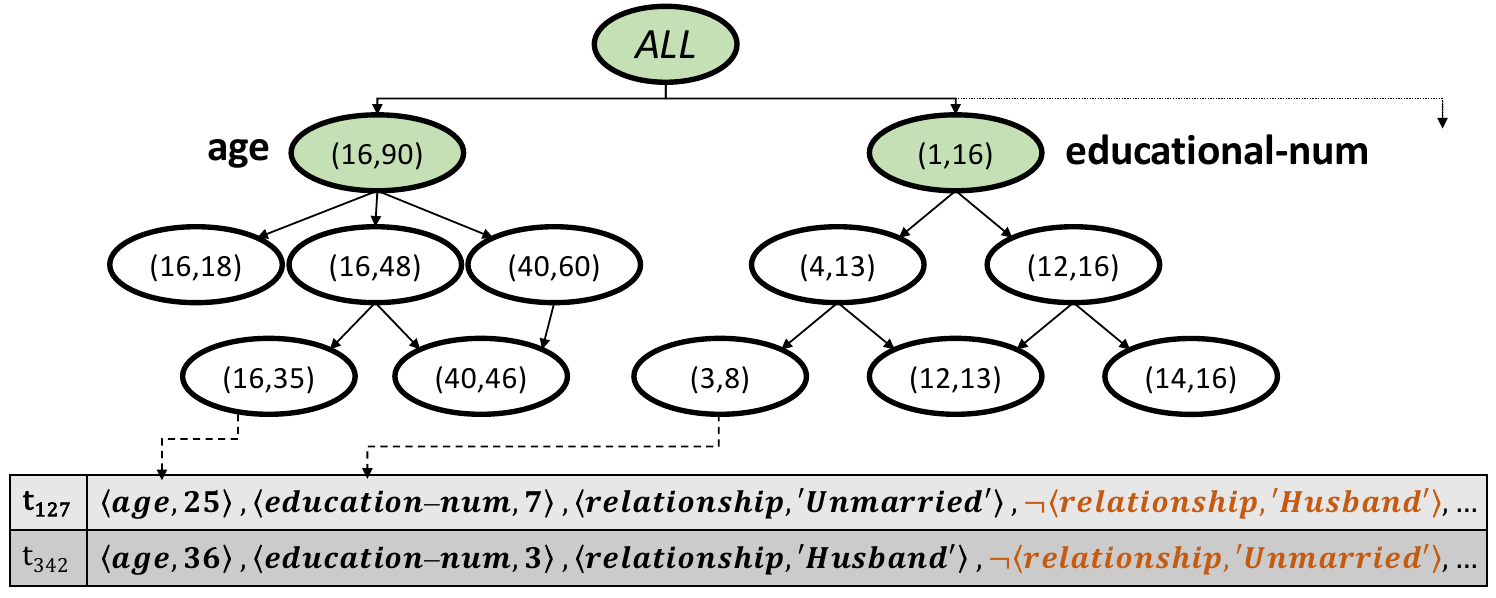}
\caption{Example augmented transactions}
\label{fig:graph_example}
\end{figure}

Recall that a gFIM algorithm operates on transactional data, where each row contains a set of discrete items associated with a categories taxonomy. A straightforward transformation from relational data to transactional form involves treating each data point $x_i$ as a set of attribute-value pairs, i.e., $t_i = \{\langle a, x_{i,a} \rangle \mid \forall a \in A\}$, where each item $\langle a, x_{i,a} \rangle$ represents an equality predicate $\langle a, =, x_{i,a} \rangle$.

However, this transactional representation is largely ineffective, as the items merely represent the raw data. This limitation restricts the ability to mine generalized patterns and, consequently, to generate effective cluster explanations.

To address this, we transform each data point into an augmented transaction that contains more generalized information. This transformation involves (1) creating an interval taxonomy for numeric values, corresponding with explanations' \textit{range} predicates and (2) injecting negations of categorical values which will represent \textit{inequality} predicates, \revb{allowing an explanation to characterize a cluster not only by what it includes but also by what it excludes.} 
Our augmentation process is described in Algorithm~\ref{alg:augment_transactions}. We next provide more details on the two segments of this process.

\SetCommentSty{rmfamily}
\IncMargin{1em}
\begin{small}
\begin{algorithm}[t]
\DontPrintSemicolon
\KwIn{Dataset \( D = \langle X, A \rangle \); numeric and categorical attribute subsets $A^N, A^C \subset A$; a set of binning methods $BIN$}
\KwOut{Augmented transactions set $T^D$, intervals taxonomy $\mathcal{T}$}

\BlankLine
$\mathcal{T} \gets$  Initialize DAG with root node $\langle ALL \rangle$; \label{ln1:tax_start}

\ForEach{numeric attribute $a \in A^N$}{
    $\mathcal{B}_a \gets \bigcup_{B \in BIN} B(X_a)$\;  
    $\alpha_M = \arg \min_{\alpha}~\{\alpha~|~[\alpha,\beta] \in \mathcal{B}_a \}$ \;
    $\beta_M = \arg \max_{\beta} \{\beta~|~[\alpha,\beta] \in \mathcal{B}_a$ \} \;
    $\mathcal{B}_a \gets \mathcal{B}_a \cup [\alpha_M,\beta_M]$ \;
    $\tau_a \gets $ Initialize a DAG with nodes $\mathcal{B}_a$ \;
    \ForEach{intervals pair $b,b' \in \mathcal{B}_a$}{
            \If{$b \prec b' \wedge \nexists b'' \in \mathcal{B}_a,~b\prec b'' \prec b'$}{
                Add edge from $b$ to $b'$ in $\tau_a$\; 
            }
        
    }
    $\mathcal{T} \gets \mathcal{T}\cup \tau_a$; add an edge from $\langle ALL \rangle$ to the root of $\tau_A$  \label{ln1:tax_end}
   
}

\BlankLine

\ForEach{data point \( x_i \in X \) }{ \label{ln1:trans_start}

    $t_i \gets \{ \langle a, x_{i,a} \rangle~|~\forall a \in A \} $ \; 
    
    \ForEach{categorical attribute $a \in A^C$}{
        \ForEach{distinct value $v \in X_a$, s.t.  $x_{i,a} \neq v $}{
            $t_i \gets t_i \cup \{\neg \langle a, v \rangle \} $\;  
        }
    }
    $T^D \gets T^D \cup \{t_i\}$\; \label{ln1:trans_end}
}

\Return{$T^D$, $\mathcal{T}$}

\caption{Generate Augmented Transactions}
\label{alg:augment_transactions}
\end{algorithm}
\end{small}

\paragraph*{Constructing the intervals taxonomy}
The purpose of the interval taxonomy is to generate effective \textit{range} predicates for numeric attributes (e.g., $Age \geq 35$ or $Education\text{-}num$ between $4\text{-}13$ years).

To achieve this, we employ multiple binning methods for each numeric attribute $a$, combining them into an interval taxonomy used as input for the gFIM algorithm. Each unique attribute-value pair $\langle a, x_{i,a} \rangle$ is 
then augmented with multiple intervals that contain it, allowing these intervals to be used as range predicates in a cluster explanation \revb{(in addition to the original values)}.

Let $A^N \subseteq A$ bet the set of numeric attributes in the dataset $D$. 
\revb{For each attribute $a \in A^N$ We first employ a predefined set of binning methods, denoted $BIN$, on $X_a$ (the values of column $a$ in $X$). Each binning method $B \in BIN$ splits $X_a$ into a set of intervals, defined as $B(X_{a}) = \{[\alpha_1,\beta_1], [\alpha_2,\beta_2], \dots\}$, s.t. $\alpha_i < \beta_i$.}

We combine multiple binning methods in our implementation of \system{}: Equal-height, Equal-width, 1-D clustering~\cite{wang2011ckmeans}, tree-based~\cite{loh2011classification}, and Optimal Binning~\cite{navas2020optimal}, \revb{and support additional methods such as domain-specific and semantic binning~\cite{setlur2022oscar}}.

Then, we construct an interval taxonomy for each attribute $a$ using the following procedure: First, we unify all binning results $ \mathcal{B}_a =  B_1(X_a) \cup B_2(X_a)\cup \dots$ and arrange them in an \textit{interval taxonomy}, built on the following strict partial order: 
\begin{equation*}
\setlength{\abovedisplayskip}{2mm}
\setlength{\belowdisplayskip}{2mm}
[\alpha_i,\beta_i] \prec [\alpha_j,\beta_j] \iff \left( \alpha_i \neq \alpha_j \lor \beta_i \neq \beta_j \right) \wedge  \left(\alpha_i \leq \alpha_j \wedge \beta_i \geq \beta_j \right)
\end{equation*}

The partial order allows as to build a semi-lattice~\cite{davey2002introduction} structure for attribute $a$, denoted $\tau_a$
that we will use for building the full taxonomy for all numeric attributes. 
$\tau_a$ is a directed acyclic graph (DAG), in which the nodes are the intervals in $\mathcal{B}_a$, and an edge from $b$ to $b'$, $b,b' \in \mathcal{B}_a$ symbolizes that $b$ is a \textit{parent} of $b'$, namely, that $b \prec b' \wedge \nexists b''~~b\prec b'' \prec b'$. 
The root of $\tau_a$ is the maximal range composed of the minimal infimum and the maximal supremum of the intervals in $\mathcal{B}_a$, i.e.  $[\alpha_M, \beta_M]$, s.t. $\alpha_M = \arg \min_{\alpha}~\{\alpha~|~[\alpha,\beta] \in \mathcal{B}_a$ \} and $\beta_M = \arg \max_{\beta} \{\beta~|~[\alpha,\beta] \in \mathcal{B}_a$ \}.
In a similar manner we construct $\tau_a$ for each numeric attribute $a \in A_N$. We then artificially combine all individual attribute taxonomies $\tau_a$ to a unified taxonomy $\mathcal{T}=\tau_{a}^1 \cup \tau_{a}^2 \cup \dots $, by creating an artificial root node labeled $All$, and creating an edge from $All$ to the root node of each $\tau_a^i$.

The final taxonomy $\mathcal{T}$ is used in the gFIM algorithm to mine \textit{generalized} frequent itemsets, each containing either a concrete item $\langle a, x_{i,a} \rangle$ or one of its ancestors in $\tau_a \subset \mathcal{T}$ (recall that item $\langle a, x_{i,a} \rangle$ is connected to all leaf nodes in $\tau_a \subseteq \mathcal{T}$ whose interval range contains the value $x_{i,a}$).

\paragraph*{Augmented transactions with value negations}
We next describe how we transform each data point $x_i$ to an augmented transaction $t_i$ (See Lines~\ref{ln1:trans_start}-\ref{ln1:trans_end} in Algorithm~\ref{alg:augment_transactions}).

For each data point $x_i$, we first convert it to a set of attribute-value pairs $\{ \langle a, x_{i,a} \rangle~|~\forall a \in A \}$. By now, items from numeric attributes are associated with their corresponding ancestors in the taxonomy $\mathcal{T}$. We then process items of categorical columns as follows: For each categorical attribute $a \in A^C$, we insert value negation items for all \textit{other} values in the column $X^a$, in a process similar to \textit{one-hot encoding}. Namely, For each $v \in X_a$, s.t. $v \neq x_{i,a}$, we add to $t_i$ the item $\neg \langle a , v \rangle $. these injected value-negation items will represent \textit{inequality predicates} in the output explanations.

For illustration, consider the following example.

\begin{example}

Figure~\ref{fig:graph_example} depicts an example of two augmented transactions with a corresponding interval taxonomy, following our running example. The transactions, $t_{124}$ and $t_{342}$, correspond to data points $x_{124}$ and $x_{342}$ in the Adult dataset, as shown in Table~\ref{tab:adults}. The augmented transactions include items for the attributes: \textit{age}, \textit{educational-num}, and \textit{relationship} (other attributes are omitted for space constraints). 
The black-colored items represent a subset of the original attribute-value pairs. Note that the numeric items are connected (using dashed lines) to corresponding leaves in the interval taxonomy, shown in the upper part of Figure~\ref{fig:graph_example}. The full taxonomy comprises two isolated sub-taxonomies, one for the attribute \textit{age} and one for \textit{educational-num} (others are omitted due to space limitations). The orange items in both $t_{124}$ and $t_{342}$ indicate value negation items for the \textit{Relationship} attribute.

After transforming the remaining data points, the gFIM algorithm is applied to the augmented transactions to generate explanation candidates, as shown in Table~\ref{tab:explanations}. These candidates are then further processed to obtain an optimal set of cluster explanations (see Figure~\ref{fig:screenshot}), as detailed below.

\end{example}

\SetCommentSty{rmfamily}
 \IncMargin{1em}
 \begin{small}
\begin{algorithm}[t]
\DontPrintSemicolon
\KwIn{Augmented Transactions set  \( T^D \); Taxonomy $\mathcal{T}$; Cluster Mapping $CL:T^D \rightarrow C$; Thresholds \( \theta_{cov}, \theta_{sep}, \theta_{con} \)}
\KwOut{A set $\optexps$ of cluster explanations for each cluster}

\( EX_{\text{all}} \gets \) Initialize results explanations dictionary\;

\ForEach{\( c \in C \)}{
    $T^D_c \gets \{t | t \in T^D \wedge CL(t)=c \}$
    
    $\mathcal{IS}_c \gets gFIM\left(T^D_c, \mathcal{T},  \text{minsup}=\theta_{cov}, 
 \text{maxsize}=\frac{1}{\theta_{con}}\right)$ \; 
    
    \( \mathcal{E}_c  \gets \{\} \) \;  
    
    \ForEach{\( IS_c \in \mathcal{IS}_c \)}{
        $E_c \gets $ Convert \( IS^c \) to a conjunction of predicates\;
        \If{\( \text{SeparationErr}(E_c) \leq \theta_{sep} \)}{
            $\mathcal{E}_c \gets \mathcal{E}_c \cup \{E_c\}$\;
        }
    }
    \( \optexps \gets \) \( SKYLINE_{E \in EX_{\theta}}(\textit{Coverage},\textit{SepError},\textit{Conciseness}) \)\;
    $EX_\text{all}[c] \gets \optexps $\;
}
\Return{\( EX_{\text{all}} \)}

\caption{Explanation Generation}
\label{alg:cluster-explorer}
\end{algorithm}
\end{small}

\subsection{Generating Explanations with gFIM}
\label{ssec:gfim}

We next describe how we mine cluster explanations using a gFIM algorithm, employed on the augmented transactions we generated as described above.  Recall that a gFIM algorithm~\cite{gfim_orig,gfim1}, as described above, takes as input a set of transactions $T = t_1, t_2,\dots$, each containing a set of discrete items, associated with some categories in an additionally provided item-category taxonomy. 
It also takes as input the maximal itemset size $maxsize$, and a \textit{support} threshold $minsup$. The support of an itemset $IS \subseteq \mathcal{I}$ is defined by: $support(IS) = \frac{|\{t | t \in T \wedge IS \subseteq t  \}|}{|T|}$.

The gFIM algorithm mines a set of frequent \textit{generalized} itemsets, $\mathcal{IS}^*$, where each resulted itemset $IS^* \in \mathcal{IS}^*$ , has~$support(IS^*) \geq minsup$ and $|IS^*| \leq maxsize$.

We next detail our explanations generation process, in which we apply a gFIM algorithm on the augmented transactions, then process the resulted itemsets into effective cluster explanations. Refer to Algorithm~\ref{alg:cluster-explorer} for pseudo code.

For each cluster $c \in C$, we first consider the transaction subsets $T^D_c$, which only contains transactions $t_i$ that are labeled with cluster $c$.
We then apply the gFIM algorithm on $T^D_c$, together with the intervals taxonomy $\mathcal{T}$, a minimal support of $minsup = \theta_{cov}$, and itemset size limit $maxsize = \frac{1}{\theta_{con}}$. 
The gFIM algorithm returns a set $\mathcal{IS}_c$ of generalized frequent items with a support value greater than $minsup$ and size under $maxsize$. 

We then further process the set $\mathcal{IS}_c$ in order to generate the optimal explanations $\optexps$ based on the following observation.

\begin{observation}
Each $IS \in \mathcal{IS}^c$ is equivalent to an explanation candidate $E_c$, having $Coverage(E_c)\geq \theta_{cov}$ and $Conciseness(E_c) \geq \theta_{con}$ 
\end{observation}

Intuitively, in each itemset $IS$, singular items of the form $\langle a, x_{i,a}$ correspond to \textit{equality predicates} of the form $\langle a, =, x_{i,a} \rangle$; 
generalized items $\langle a, [\alpha, \beta] $ are equivalent \textit{range} predicates of the form $\langle a,\text{between}, [\alpha, \beta] \rangle$; and categorical value negations of the form $\neg \langle a , v \rangle$ are \textit{inequality} predicates, i.e., $\langle a, \neq, v  \rangle $. 

Naturally, the coverage of $E_c$ is higher than $\theta_{cov}$, since the support of $IS^c$ is guaranteed (by the gFIM algorithm) to be higher than $\theta_{cov}$. Similarly, the conciseness score $E_c$ is higher than $\theta_{con}$, since we limit the maximal size of the itemset using $\frac{1}{\theta_{con}}$.  

Let $\mathcal{E}^c$ be the candidate explanations generated from transforming the gFIM output $\mathcal{IS}^c$ to a conjunction of predicates as explained above. Then, in order too obtain the subset $\optexps \subseteq \mathcal{E}^c $ of optimal explanations for cluster $c$, we only need to filter the candidate explanations using the separation error threshold $\theta_{sep}$ then find the Pareto optimal explanations , w.r.t. coverage, separation error and conciseness. The latter step is done using the skyline operator~\cite{borzsony2001skyline} on the filtered candidate explanations set, $SKYLINE_{E \in \mathcal{E}^c}(\textit{Coverage},\textit{SepError},\textit{Conciseness})$.

The results of the skyline operator retrieves the final desired set of explanations $\optexps$ (see an example explanation for Cluster 0 and Cluster 1 from our running example  in Figure~\ref{fig:screenshot}).

\paragraph*{Cost Discussion}
The overall computational cost associated with the \system{} explanation generation process primarily aligns with the cost of the gFIM algorithm. The preprocessing phase (transaction augmentation) and the postprocessing phase (conversion back to predicates and skyline computation) exhibit linear or small polynomial costs, whereas the cost of \textit{gFIM} can be exponential in the number of items (+ generalized items)~\cite{gfim_orig,gfim2}.

\system{} effectively neutralizes the potentially high costs of gFIM by naturally restricting the number of considered items. This restriction arises from the relatively high minimum support ($minsup$) threshold used, which is aligned with the targeted explanation coverage threshold $\theta_{cov}$. Unlike conventional \textit{gFIM} applications, where users might choose low support thresholds (e.g., 0.01 to 0.05), such settings are impractical in our context due to the need for explanations that adequately cover \textit{the majority} of data points within a cluster.

To further reduce the number of items processed and thereby lower computational costs, we introduce an \textit{attribute selection} technique used in \system{}. By selectively limiting computation to promising attributes, the gFIM algorithm operates solely on items associated with these attributes, and therefore obtains an additional, significant speedup.

\subsection{Attribute Selection Optimization} 
\label{ssec:sampling}

\begin{table}[t]
\centering
\small
\begin{tabular}{lcc}
\hline
\textbf{Dataset Name}                                  & \textbf{\#Rows} & \textbf{\# Attr.} \\ \hline

Urban Land Cover & 168 & 148 \\
DARWIN   & 174  & 451  \\
Wine  & 178   & 13 \\
Flags & 194 & 30  \\
Parkinson Speech
 & 1040 & 26 \\
Communities and Crime & 1994 & 128 \\
Turkiye Student Evaluation  & 5820 & 33  \\
in-vehicle coupon recommendation  & 12684 & 23 \\
Human Activity Recognition& 10299 & 561\\
Quality Assessment of Digital Colposcopies & 30000 & 23 \\
RT-IoT2022 & 123117 & 85 \\
Gender by Name & 147270 & 4 \\
Multivariate Gait Data & 181800 & 7 \\
Wave Energy Converters & 288000 & 49 \\
3D Road Network & 434874 & 4 \\
Year Prediction MSD & 515345 & 90 \\
Online Retail & 1067371 & 8 \\
MetroPT-3 Dataset & 1516948 & 15 \\
Taxi Trajectory & 1710670 & 9 \\
\hline
\end{tabular}
\caption{List of UCI Datasets~\cite{uci_datasets} and their properties }
\label{tab:dataset_summary}
\end{table}

One of the most crucial dataset properties affecting frequent itemset mining on relational data is the number of attributes, as the number of attributes directly influences the number of items (key-value pairs)~\cite{fu1995meta}. 
To mitigate this effect, \system{} introduces a simple yet effective attribute selection technique based on feature importance calculation~\cite{zien2009feature}, where we focus on the most pivotal attributes for each cluster \( c \in C \).

Our attribute selection method is applied once for a dataset \( D \) and used for generating explanations for all clusters in \( C \). 

For each cluster \( c \in C \), we train a binary decision tree classifier \( M_c(X, \text{CL}(x)=c) \), i.e., predicting if a data point is labeled as \( c \) (true) or not (false). We define the maximal depth to be \( \frac{1}{\theta_{con}} \). 
To obtain an importance score for each attribute \( a \in A \), we calculate the mean Gini impurity~\cite{breiman1984classification} score (known as Gini importance) \( G_c(a) \), and select the top \( n_{\text{attr}} \) attributes. 
We then average the Gini scores across all models \( M_c \) to obtain the final importance score \( \text{attr-score}(a) = \frac{\sum_{c \in C} G_c(a)}{|C|} \).
Finally, we select the top \( n_{\text{attr}} \) features obtaining the highest attr-score.

Note that our importance measure assigns equal weights to all clusters, thus avoiding bias towards larger clusters that may hinder the explanation quality for smaller ones. In our implementation of \system{}, we set $n_{attr}$ to be proportional to the conciseness threshold $\theta_{con}$, using a scaling parameter $p$: $n_{\text{attr}} = \lfloor\frac{1}{\theta_{con}} \times p\rfloor$. 

As detailed in Section~\ref{sec:experiments}, we experiment with \( 1 \leq p \leq 2 \) and compare the results to the exact computation, where \( n_{\text{attr}} = |A| \). Our results are highly positive, showing that our attribute-selection optimization allows for an average speedup of 14.4X, and may reach up to 26X or higher for larger datasets, while negligibly affecting the quality of the explanations (see Section~\ref{sec:exp_results}).

%% file: experiments.tex
\section{experiments}\label{sec:experiments}

We evaluated \system{} on 98 clustering results, comparing its explanation quality and runtime to several XAI baselines, \revall{with an additional user study to validate the quality assessment}. Our results shows that \system{} produces superior explanations, with a 12X better running times than the closest baseline. Further examining the effectiveness of attribute selection, we show it achieved a 14.4X speedup with minimal impact on quality.




\begin{table}[t]
\centering
\small
\begin{tabular}{|l|l|l|l|l|}
\hline
\textbf{PID} & \textbf{Scaling} & \textbf{OneHot} & \textbf{PCA} & \textbf{Clustering Algorithm} \\ \hline
1 & \cmark & \cmark & \cmark & K-Means \\ \hline
2 & \cmark & \cmark & \xmark & K-Means \\ \hline
3 & \xmark & \cmark & \cmark & K-Means \\ \hline
4 & \xmark & \cmark & \xmark & K-Means \\ \hline
5 & \cmark & \cmark & \cmark & DBSCAN \\ \hline
6 & \cmark & \cmark & \xmark & DBSCAN \\ \hline
7 & \xmark & \cmark & \cmark & DBSCAN \\ \hline
8 & \xmark & \cmark & \xmark & DBSCAN \\ \hline
9 & \cmark & \cmark & \cmark & Birch \\ \hline
10 & \cmark & \cmark & \xmark & Birch \\ \hline
11 & \xmark & \cmark & \cmark & Birch \\ \hline
12 & \xmark & \cmark & \xmark & Birch \\ \hline
13 & \cmark & \cmark & \cmark & Spectral \\ \hline
14 & \xmark & \cmark & \cmark & Spectral \\ \hline
15 & \cmark & \cmark & \cmark & Affinity Propagation \\ \hline
16 & \xmark & \cmark & \cmark & Affinity Propagation \\ \hline
\end{tabular}
\caption{Clustering pipelines used in our experiments}
\label{tab:pipelines_overview_final}
\end{table}

\subsection{Experimental Setup}
We next describe our benchmark dataset of clustering results, the explanation quality measures we use, the prototype implementation of \system{}, and the baselines approaches.


\paragraph*{Benchmark Dataset}
To evaluate \system{}, we developed a benchmark dataset containing 98 clustering results obtained from 16 different clustering pipelines using 5 clustering algorithms, applied to 19 datasets.
We used publicly available dataset from the UCI collection~\cite{uci_datasets}, chosen to ensure a wide range of data shapes: Lengths between 168 and 1.7M rows, and width between 4 and 561 columns. See Table~\ref{tab:dataset_summary} for exact details.

Table~\ref{tab:pipelines_overview_final} details the exact steps used in each pipeline, each comprises a combination of the following steps: (1) Standard scaling for numeric columns, (2) One-hot encoding for categorical data, (3) Dimensionality reduction using PCA~\cite{dunteman1989principal} to a number equal to 90\% of the original columns, and (4) Applying one of five common clustering algorithms: K-Means~\cite{kmeans_paper}, DBSCAN~\cite{dbscan_paper}, Birch~\cite{zhang1996birch}, Spectral Clustering~\cite{spectral_paper}, and Affinity Propagation~\cite{affinity_propagation}. 
All clustering pipelines were executed over all 19 dataset, returning for each dataset $D=\langle X, A\rangle$ a clustering label function $CL(x)$ for the data points $X$. 
Finally, we filtered out unsuccessful clustering results where the distance between in-cluster data points is similar to that of different-cluster data points, to avoid explaining close-to-random results. This was implemented by calculating the \textit{silhouette coefficient}~\cite{rousseeuw1987silhouettes} $SC(X, CL(x))$ and filtering out clustering results with low scores, below 0.1.
The final benchmark collection includes 98 clustering results instances.

\paragraph*{Evaluating Metrics} 
Based on the clustering explanations quality criteria described in Section~\ref{ssec:measures}, we use a unified Quality Score for an Explanation (QSE) which balances the coverage, separation error, and conciseness. QSE is defined for a \textit{single} cluster explanation by:    
\begin{equation*}
\setlength{\abovedisplayskip}{2mm}
\setlength{\belowdisplayskip}{2mm}
\small
    \text{QSE}(E_c) = \frac{\textit{Coverage}(E_c) + \left(1 - \textit{SeparationErr}(E_c)\right) + \textit{Conciseness}(E_c)}{3}
\end{equation*}

Recall that \system{} (as well as the baselines introduced below) may produce multiple explanations for each cluster. Given the results of a clustering pipeline, $CL(x)$, we use each baseline to generate a set of explanations for each cluster. Let $EX_{all} = \{ \mathcal{E}_c \mid \forall c \in C \}$ be the set of all explanation sets produced by the baseline for a given clustering results instance. We then aggregate the QSE scores for $EX_{all}$ by calculating the average score of the best explanation generated for each cluster:

$$QSE(EX_{all})  \frac{\sum_{\mathcal{E}_c \in EX_{all}} \max_{E_c \in \mathcal{E}_c}QSE(E_c)}{|C|}$$

We also measure the execution time of each baseline, as the time it takes to generate \textit{all} cluster explanations $EX_{ALL}$.

\paragraph*{\system{} Implementation \& Configuration}
\sys\ is implemented in Python 3.10. It uses Pandas \cite{reback2020pandas} to store and manipulate the dataset and NumPy \cite{harris2020array} and calculate gFIM via the Pythonic implementation for Apriori\footnote{\url{https://efficient-apriori.readthedocs.io/en/latest/}}. We have made the source code available in~\cite{our_github}. 
The experiments were conducted on a Windows 10 laptop with 32GB RAM and 3.6 GHz CPU.

As for the \system{} configuration, we used the attribute selection method (as described in Section~\ref{ssec:sampling}), using $p=1$, for both quality and running times experiments. 
The thresholds configuration used is:
coverage threshold \( \theta_{cov}=0.8 \); separation error threshold \( \theta_{sep} =0.3 \); and conciseness threshold \( \theta_{con}=0.2 \), allowing for a maximum of 5 predicates in each explanation. 
These thresholds, when possible, were used for the baseline approaches as described below.

\paragraph*{Baseline Approaches} 

We compared \system{} to several XAI baselines~\cite{lundberg2017unified,ribeiro2018anchors,du2019techniques,friedman2008predictive,cohen1999simple}. For the first two, we fitted an auxiliary XGBoost~\cite{chen2016xgboost} model to predict clustering labels, then used the baselines to explain its results. The third baseline followed a similar approach but used a Random Forest internally. The fourth baseline uses a simpler auxiliary model -- a standard decision tree.




\begin{figure*}[t]
\vspace{-3mm}
    \centering
    \begin{subfigure}[b]{0.32\textwidth}
        \includegraphics[width=\textwidth]{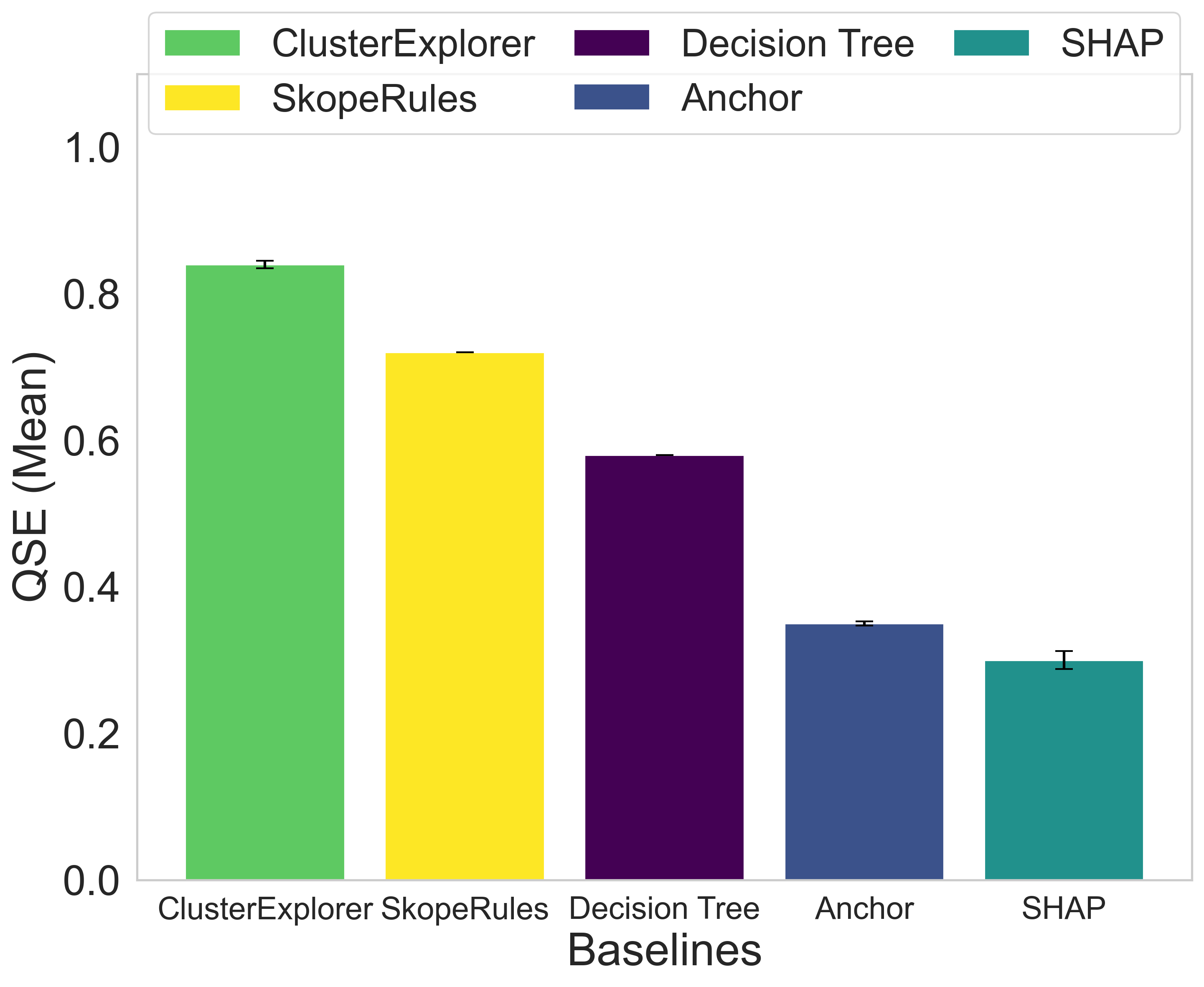}
        \caption{\reva{Average QSE}}
        \label{fig:sum_mean_scores}
    \end{subfigure}
    \hfill
    \begin{subfigure}[b]{0.32\textwidth}
        \includegraphics[width=\textwidth]{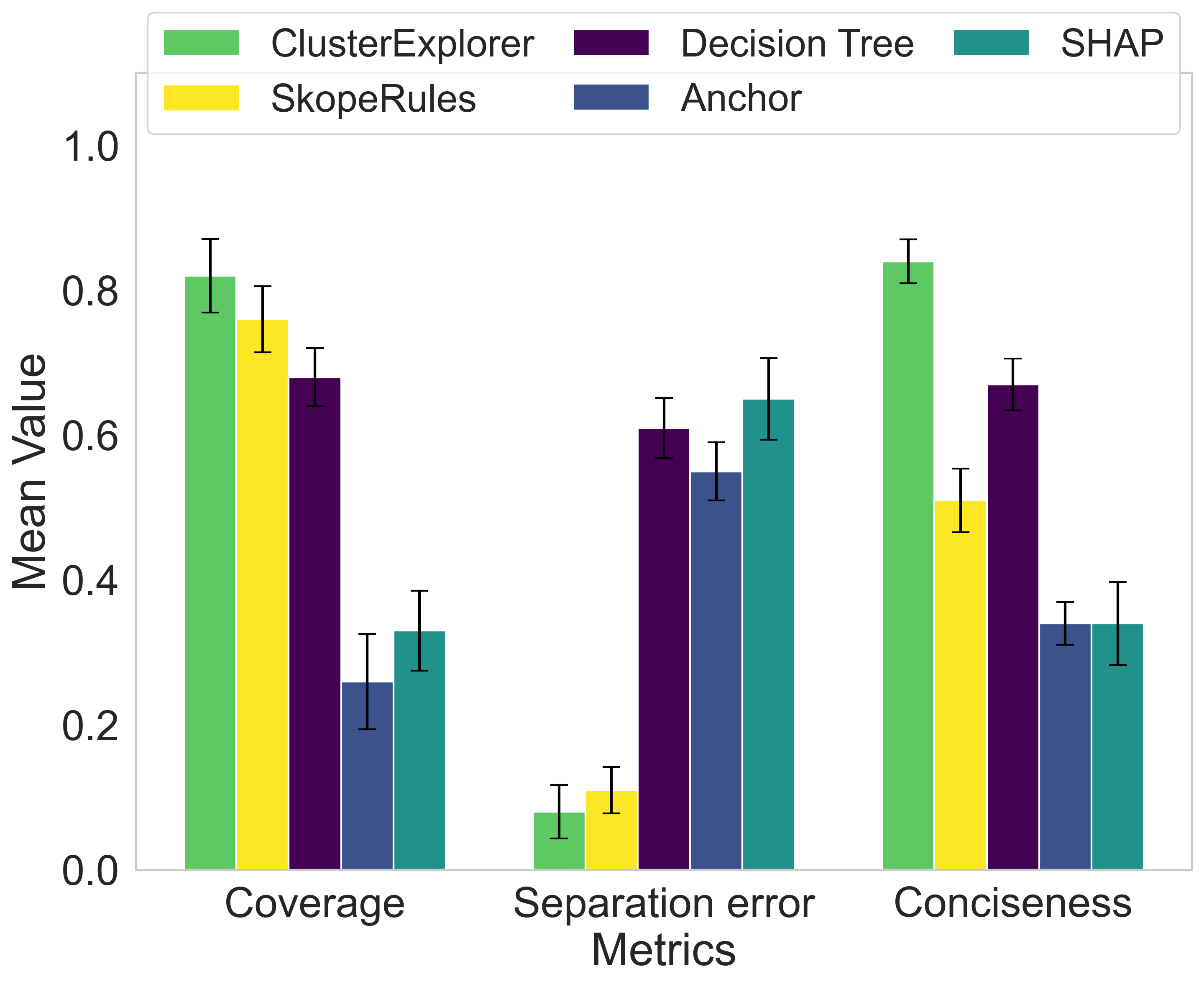}
        \caption{\revall{Average Score for each metric}}
        \label{fig:mean_metrics}
    \end{subfigure}
        \hfill
    \begin{subfigure}[b]{0.32\textwidth}
        \includegraphics[width=\textwidth]{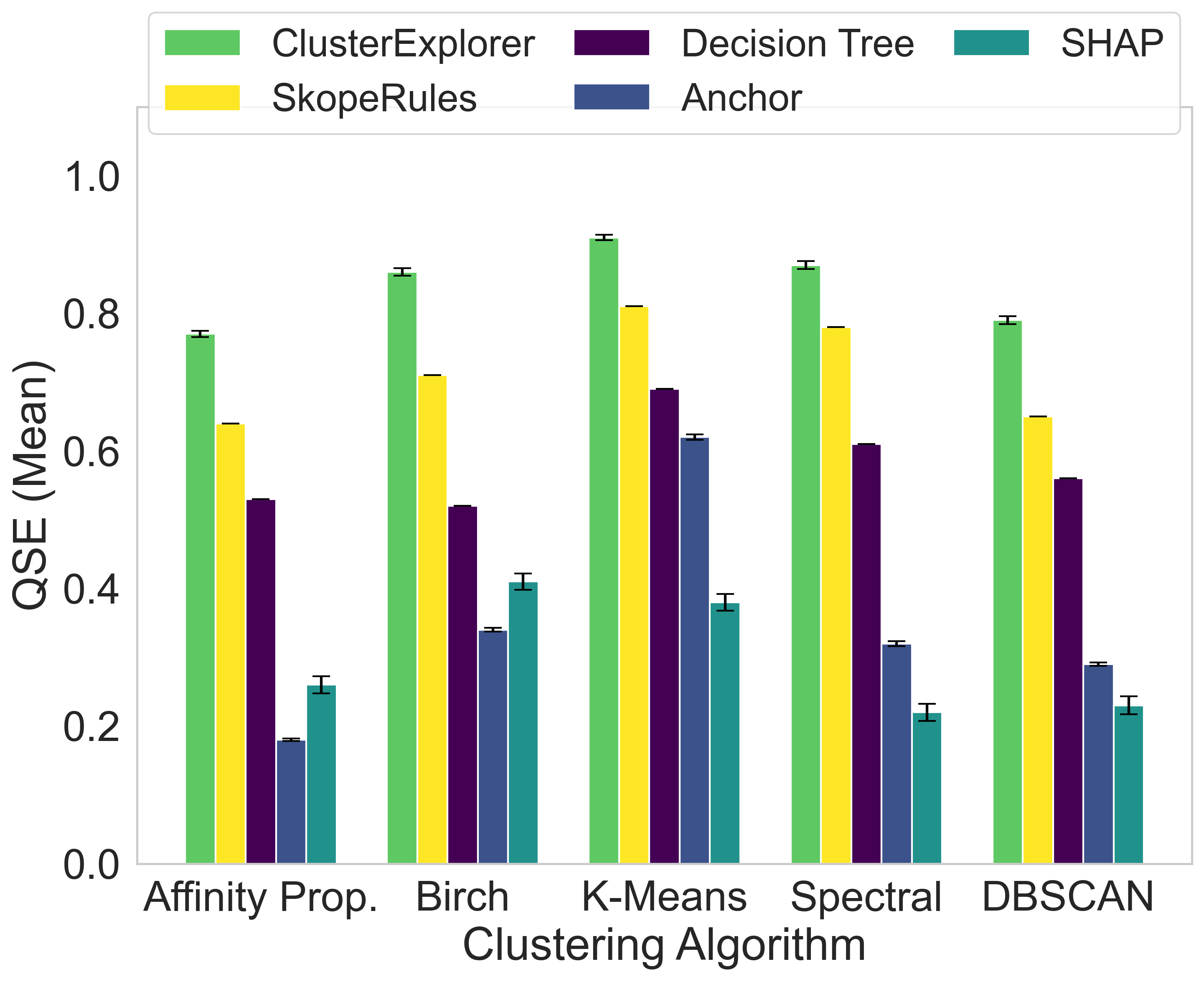}
        \caption{\reva{Average QSE per Clustering Algorithm}}
        \label{fig:qse_clustering}
    \end{subfigure}

    \caption{\revall{Combined figures illustrating QSE and metrics analysis.}}
    \label{fig:combined_figures}
\end{figure*}

We next describe the baselines in more detail. 

\noindent {\bf 1. SHAP} \cite{lundberg2017unified}: SHAP is a highly popular game-theoretic approach for explaining individual predictions of machine learning models. SHAP provides an importance score for each attribute-value pair of a given data point $x$ and the model prediction $M(x)$.

To utilize SHAP for deriving cluster explanations, we first employ XGBoost~\cite{chen2016xgboost} to fit the clustering labels $CL(X)$. Then, to produce explanations for each individual cluster, we calculate the SHAP scores of all correct predictions for a given cluster, i.e., where $M(x)=c$ (we limited the number of samples to 2000 due to expensive execution costs). Since SHAP scores are computed for each attribute-value pair, we applied equal-width binning to each numeric attribute and calculated the average SHAP value for each bin. We returned as explanations the conjunction of the top-$i$ bins (or categorical key-value pairs), for $1 \leq i \leq \frac{1}{\theta_{con}}$.

\noindent {\bf 2. Anchors} \cite{ribeiro2018anchors}: The Anchors framework is a more recent solution for local explanations, aiming to provide more concise and meaningful explanations than previous solutions such as SHAP~\cite{lundberg2017unified} and LIME~\cite{ribeiro2016should}, which provide an importance score for each feature-value combination. In Anchors, the explanations are provided as decision rules that ``anchor'' the prediction, i.e., changes made to other attributes or ranges that do not appear in those rules do not affect the model outcome.
For this baseline, we also utilize an XGBoost model to fit the cluster labels and then produce the anchors (decision rules) for the data points of each cluster. Since rules are produced for individual data points, we return a sample of 20 such explanations for each cluster.

\noindent {\bf 3. SkopeRules\footnote{https://github.com/scikit-learn-contrib/skope-rules}}: SkopeRules is an open-source library for generating decision rules that characterize a target class, specifically identifying class instances with high precision. The SkopeRules code is based on a line of previous research for extracting decision rules from ensemble learning models~\cite{friedman2008predictive,cohen1999simple}. According to their documentation, SkopeRules utilizes a Random Forest model to fit the class labels and then mines a diverse set of rules from the individual decision trees generated by the model. The program allows setting the trees' depth, thereby limiting the explanations' size, as well as thresholds for precision and recall (we used $\frac{1}{\theta_{con}}$, $\theta_{cov}$, and $1-\theta_{sep}$ for those thresholds).

\noindent {\bf 4. Decision-Tree Explanations}:
Last, inspired by the notion of ML \textit{interpretability}~\cite{du2019techniques}, we devise an additional baseline based on fitting a simple decision tree model to the cluster labels. Unlike the previous baselines, which utilize a complex tree ensemble model (XGBoost) coupled with an external explanation framework, the decision tree model is simple and easy to interpret.

We derive cluster explanations by first fitting a binary decision tree with a maximum depth of $\frac{1}{\theta_{con}}$, then returning all tree paths from root to leaf as cluster explanations.

\subsection{Experiment Results}
\label{sec:exp_results}
We next detail our results, examining first the quality of generated explanations, then the running times comparison, and finally, the effect of our attribute-selection optimization on both the explanation quality and running times.

\subsubsection{Explanation Quality Evaluation (Automatic)}
\label{ssec:quality}

Figure~\ref{fig:sum_mean_scores} depicts the explanation set QSE score for each baseline, averaged across all 98 clustering result instances \reva{along with a vertical error bar reporting the .95 confidence interval.}
First, we note that the two baselines based on local explanations for supervised models—SHAP and Anchors—both demonstrate inferior results, with scores of 0.32 and 0.35, respectively. Decision tree explanations, based on the interpretability-by-design approach, achieve a higher average QSE of 0.58, surpassing SHAP and Anchors but still significantly lower than SkopeRules and \system{}. SkopeRules, which mines rules from a more complex Random Forest model, achieves higher QSE scores (0.72) than the simple decision tree but is still 13 points lower than \system{}, which attains the highest average QSE of 0.84.

\revall{In Figure~\ref{fig:mean_metrics} we further report the average maximal score obtain for each metric individually. 
\system{} achieves the highest scores across all individual metrics. Comparing baselines, see, for example, that SkopeRules shows a comparable separation error (0.11, only 0.03 higher than \system{})  -- but falls behind in coverage (0.76 vs. 0.82) and conciseness (0.51 vs. 0.84),  whereas Decision Tree explanations are rather concise (0.67) with fair coverage (0.68) but suffer from a much higher separation error (0.61). 
}

\revb{
Next, we analyze the distribution of QSE scores across different clustering algorithms, as shown in Figure~\ref{fig:qse_clustering}. \system{} again outperforms the other baselines, but, interestingly, the results slightly vary: The highest performance is achieved for K-Means clusters (0.91), while the lowest is observed for Affinity Propagation (0.77). This variation can be intuitively attributed to the differences in cluster shapes: K-Means generates spherical clusters centered around distinct points, whereas Affinity Propagation, which employs a message-passing approach~\cite{affinity_propagation}, produces clusters with more complex and irregular shapes. Similar differences in QSE scores are observed for the additional baselines as well. 
}

\subsubsection{\revall{User Study}}
\label{ssec:userstudy}
\revall{
To further validate the quality of the explanations, we conducted a small-scale user study. We recruited 12 participants by public calls targeting computer science students or graduates familiar with data analysis and/or data science. For the datasets \textit{Wine}, \textit{Turkiye student evaluation},
\textit{Communities and Crime}, we randomly selected 12 clustering pipeline results for evaluation. We chose these datasets due to their moderate complexity, ensuring participants could understand the data and its attributes without requiring additional time or effort.
}

\begin{figure}[t]
    \includegraphics[width=0.7\columnwidth]{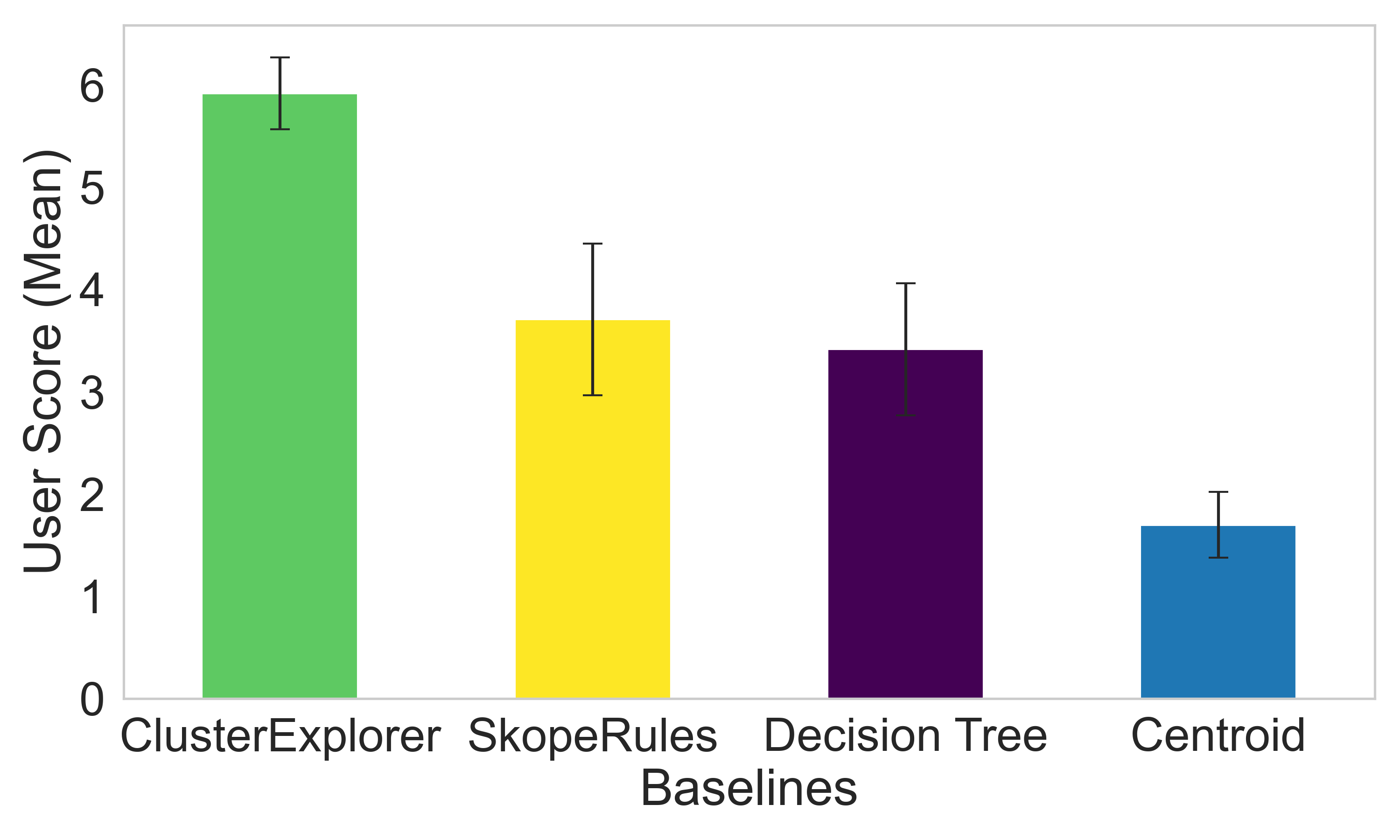}    
    \caption{\revall{User Study Results}}
    \label{fig:user_study}
\end{figure}

\begin{figure*}[t]
\vspace{-3mm}
    \centering
    \begin{subfigure}[b]{.32\textwidth}
        \centering
        \includegraphics[width=\linewidth]{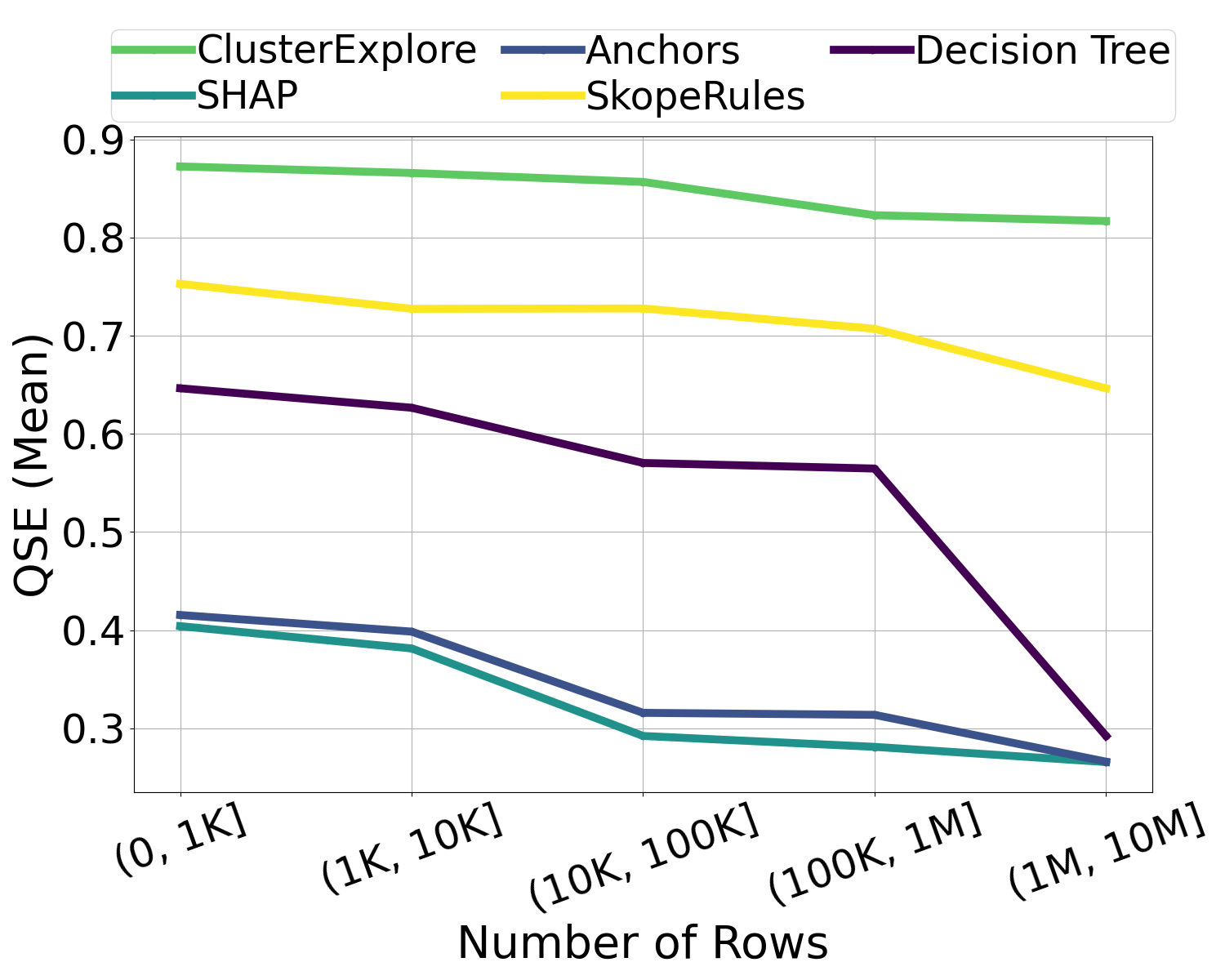}
        \caption{QSE vs. Number of Rows}
        \label{fig:sum_mean_vs_rows}
    \end{subfigure}%
    \begin{subfigure}[b]{.32\textwidth}
        \centering
        \includegraphics[width=\linewidth]{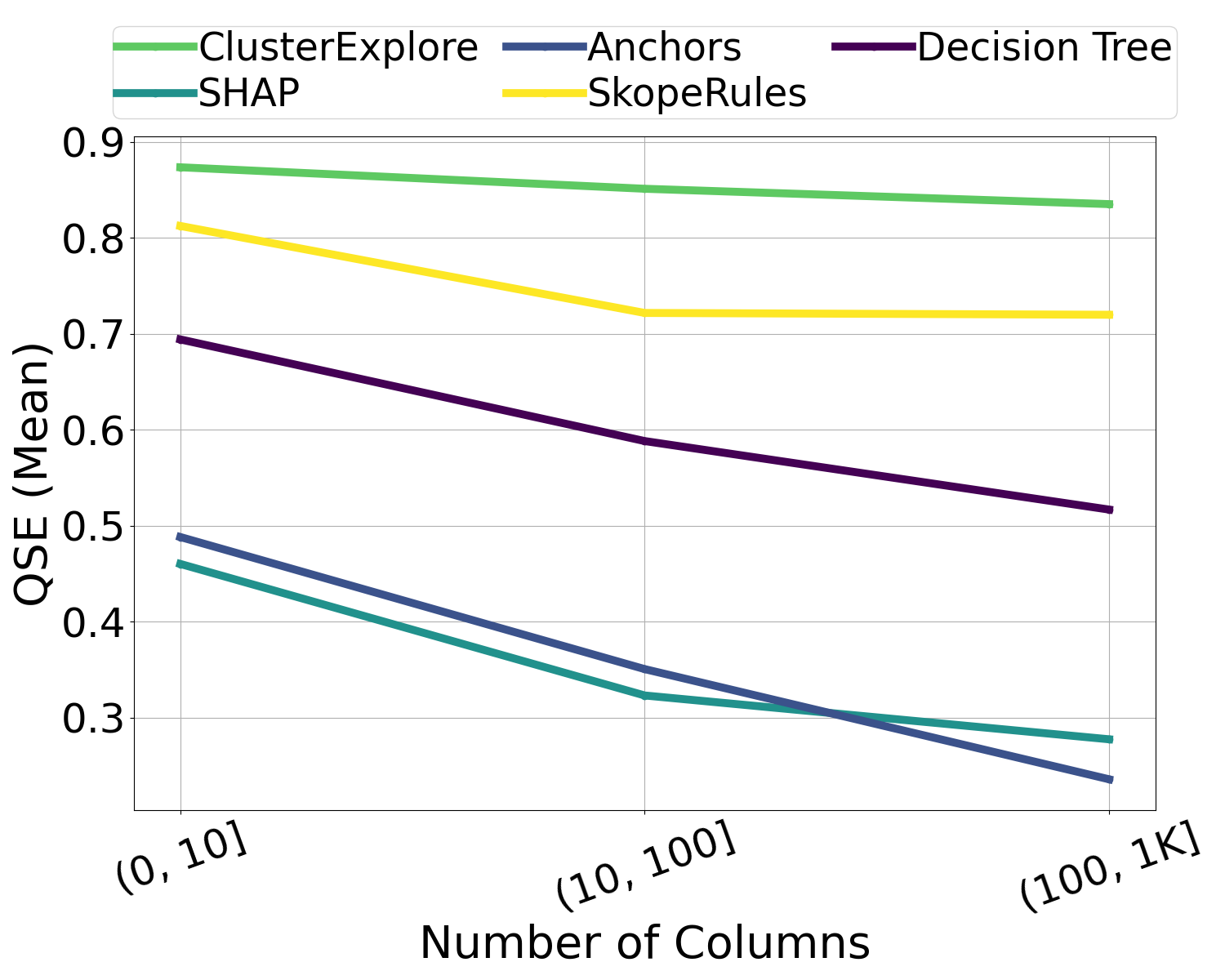}
        \caption{QSE vs. Number of Columns} 
        \label{fig:sum_mean_vs_columns}
    \end{subfigure}%
    \begin{subfigure}[b]{.32\textwidth}
        \centering
        \includegraphics[width=\linewidth]{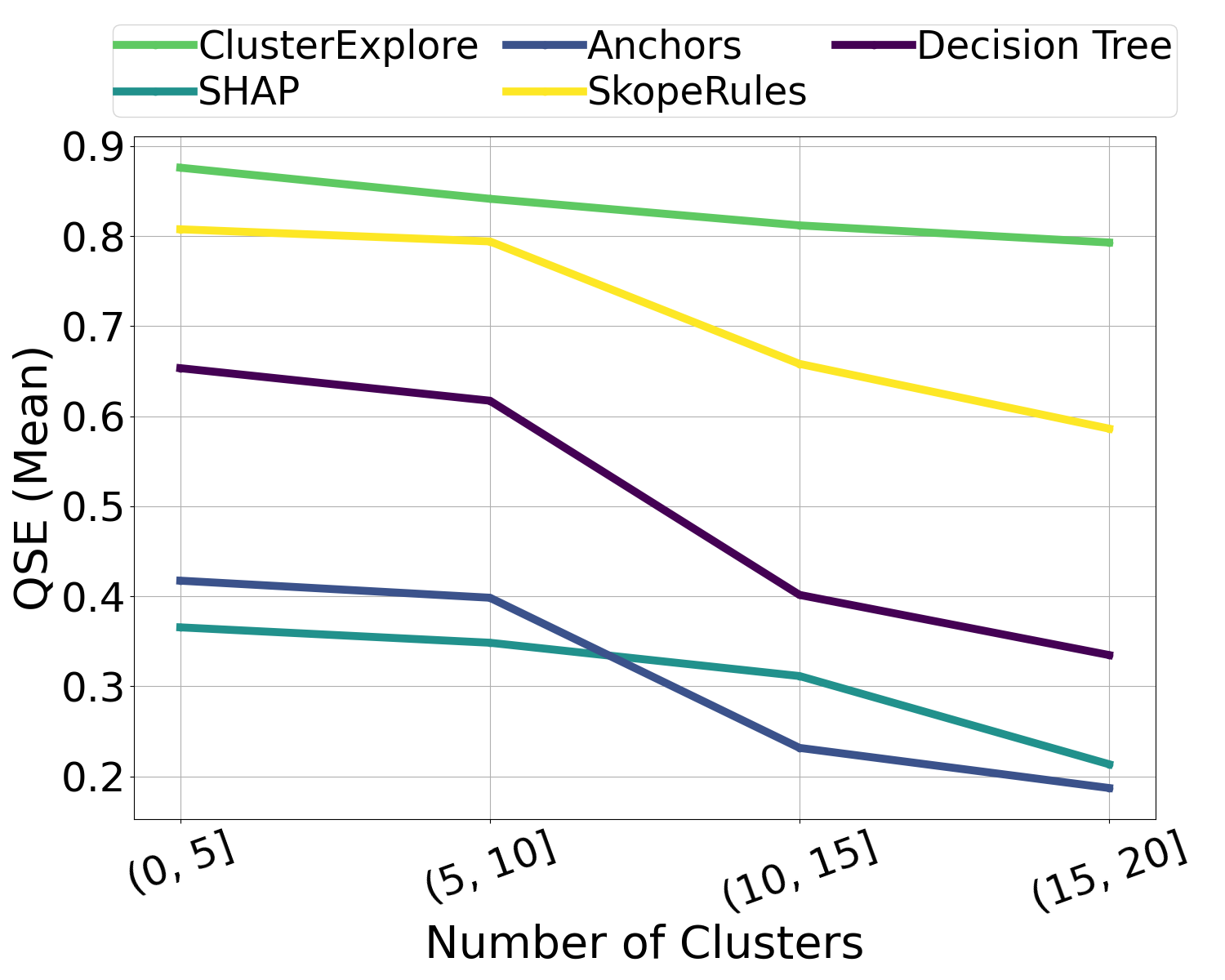}
        \caption{QSE vs. Number of Clusters}
        \label{fig:sum_mean_vs_clusters}
    \end{subfigure}
    \caption{Average QSE scores as a factor of the number of rows, columns, and clusters}
    \label{fig:qse_lines}
\end{figure*}

\revall{Each participant was first introduced to the data and its attributes, followed by a visualization of the clustering pipeline results in a two-dimensional plot (as in Figure~\ref{fig:clusters_vis}). Subsequently, we presented a series of cluster explanations generated by the following methods: \system{}, SkopeRules, Decision-Tree, and an alternative non-rule-based approach using \textit{cluster centroids} (i.e., the data point closest to the cluster mean). We excluded Anchor and SHAP  as their explanations consistently yielded significantly lower QSE scores as described above. 
For all rule-based approaches (\system{}, SkopeRules, and Decision-Tree), we additionally described the explanation \textit{coverage} and \textit{separation error} rates (phrased as in Figure~\ref{fig:screenshot}) to provide further context for evaluation. Participants were then asked to rate the quality of each explanation on a scale from 1 (low) to 7 (high) without knowing the associated method used to generate the explanations. They were also required to justify their ratings through free-text responses.
Each participant evaluated three cluster explanations per dataset, resulting in a total of 36 user evaluations, equally distributed over the 12 clustering results.
}

\revall{
The average user scores are presented in Figure~\ref{fig:user_study} (with vertical lines depicting .95 CI). Participants unanimously preferred rule-based explanations, with cluster centroids receiving the lowest average score of 1.7. Notably, the user scores ranking is strongly correlated with the average QSE scores: \system{} achieved the highest average score of 5.92, compared to 3.71 for SkopeRules and 3.42 for Decision Tree.

Analyzing the participants' justification for their ratings, \system{} was consistently rated as the best approach due to its combination of high coverage, accuracy, and diversity of explanations. SkopeRules was appreciated for its accuracy (i.e., low separation error) but was criticized for limited coverage and lack of explanation variety. Decision Tree was noted for its concise explanations but was penalized for significantly higher error rates. Explanations based on centroids were generally considered overly complex and not user-friendly, compared to the rule-based approaches.

}

\subsubsection{parameters effect on explanations quality}
\label{ssec:param_effect}
We further investigate the effect of data and problem complexity on the QSE score.

Figure~\ref{fig:qse_lines} shows average QSE scores as a function of the number of rows, columns, and clusters. The decision tree baseline is most affected, with QSE scores dropping by 54.8\%, 25.5\%, and 48.7\% as rows, columns, and clusters increase, respectively, reflecting its limitations for large, complex datasets~\cite{arrieta2020explainable,doshi2017towards}. Other baselines also degrade with data complexity: SkopeRules drops by 14.1\%, 11.4\%, and 27.4\%, Anchors by 35.9\%, 51.7\%, and 55.2\%, and SHAP by 34.2\%, 39.7\%, and 41.6\% for datasets with over 1M rows, 100 columns, and 15 clusters, respectively. In contrast, \system{} remains stable, with minor decreases of 6.3\%, 4.4\%, and 9.4\%.

\begin{figure}[t]
    \includegraphics[width=0.95\columnwidth]{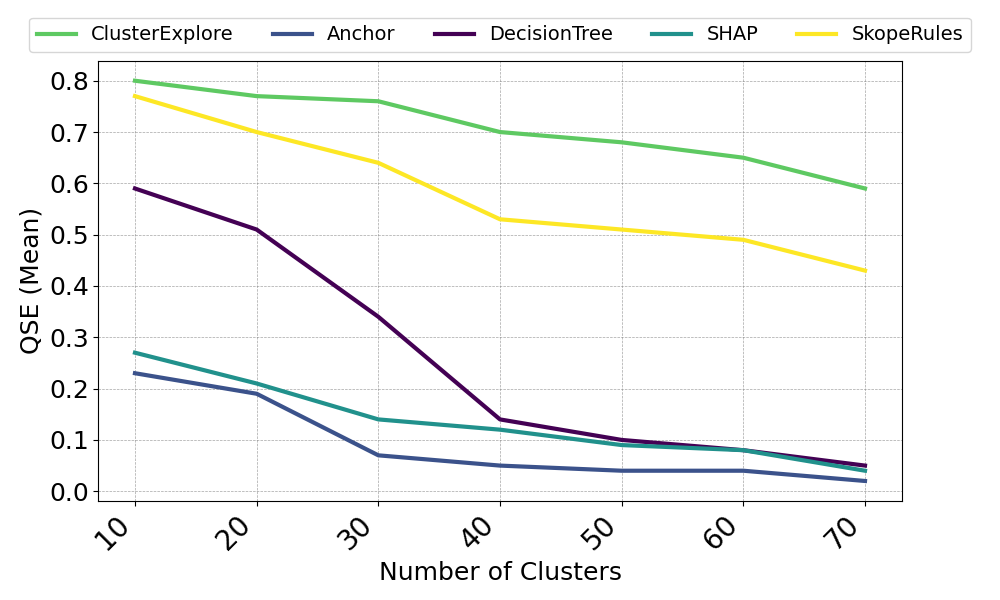}    
    \caption{\revc{QSE vs. Num. of Clusters(Simulated)} }
    \label{fig:More_clusters_QSE}
\end{figure}

\revc{
Finally, to take a closer look at the performance as a function of the number of clusters, we conducted additional experiments on simulated datasets. In our benchmark of real datasets, all pipelines producing more than 20 clusters yielded Silhouette scores below the acceptable threshold ($\leq 0.1$). To address this, we generated artificial  datasets using the method from~\cite{pedregosa2011scikit}, each with 10K rows, 10 columns, and between 10 to 70 centroids. We then executed all clustering pipelines (see Table~\ref{tab:pipelines_overview_final}), resulting in 5490 clustering results with Silhouette scores $\geq 0.1$. Figure~\ref{fig:More_clusters_QSE} illustrates the average QSE of the baselines as a function of the number of clusters. While all baselines show a decline in quality as the number of clusters increases, \system{} exhibits the most moderate decline, achieving a mean QSE of 0.59 for the cases of 70 clusters, compared to 0.43 for SkopeRules and less than 0.05 for the remaining baselines.
}

\begin{figure*}[t]
\vspace{-1mm}
    \centering
    \begin{subfigure}[b]{.32\linewidth}
     \includegraphics[width=\linewidth]{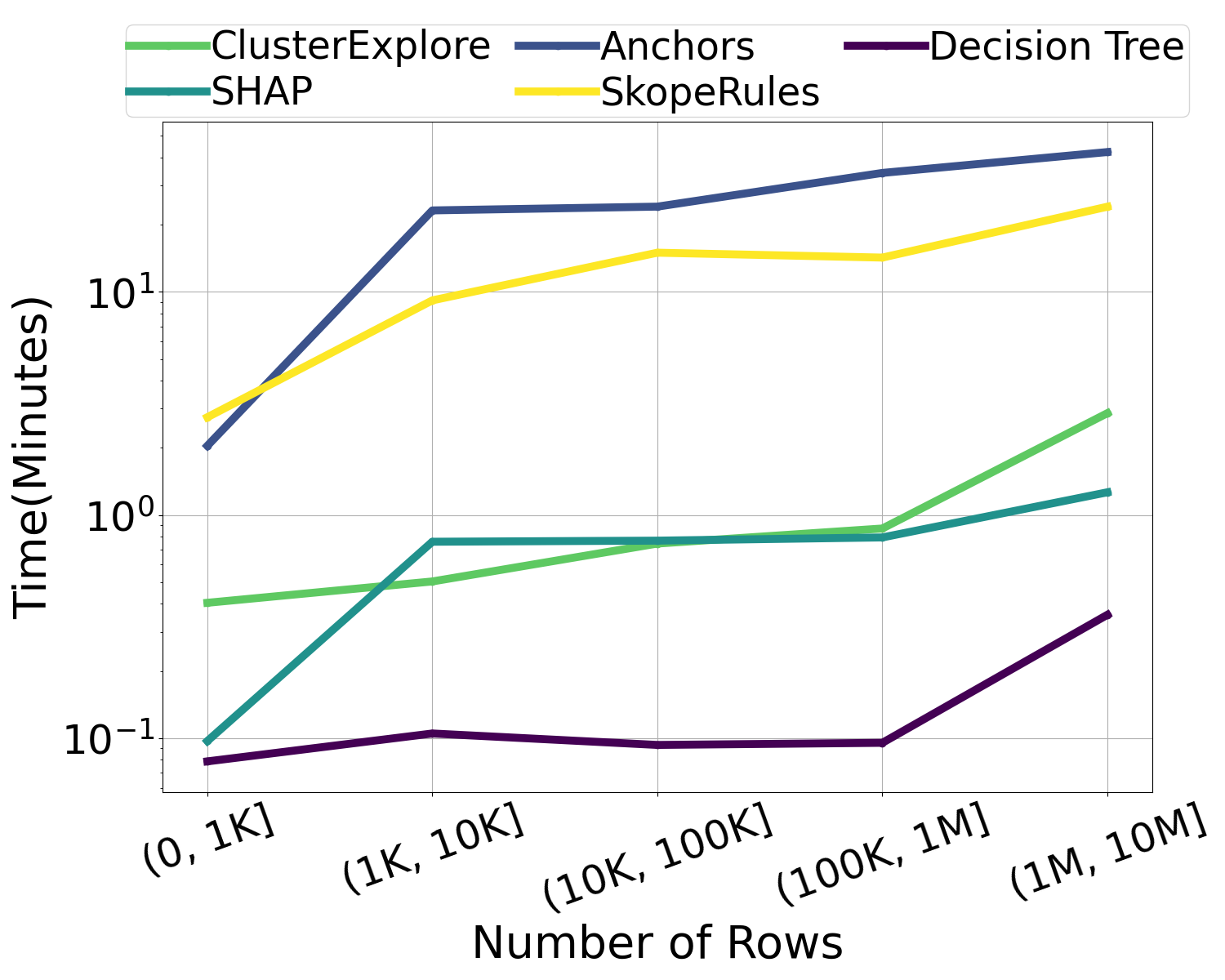}
    \caption{Runtime vs. Number of Rows}
    \label{fig:time_vs_rows}
    \end{subfigure}%
    \begin{subfigure}[b]{.32\linewidth}
    \centering
     \includegraphics[width=\linewidth]{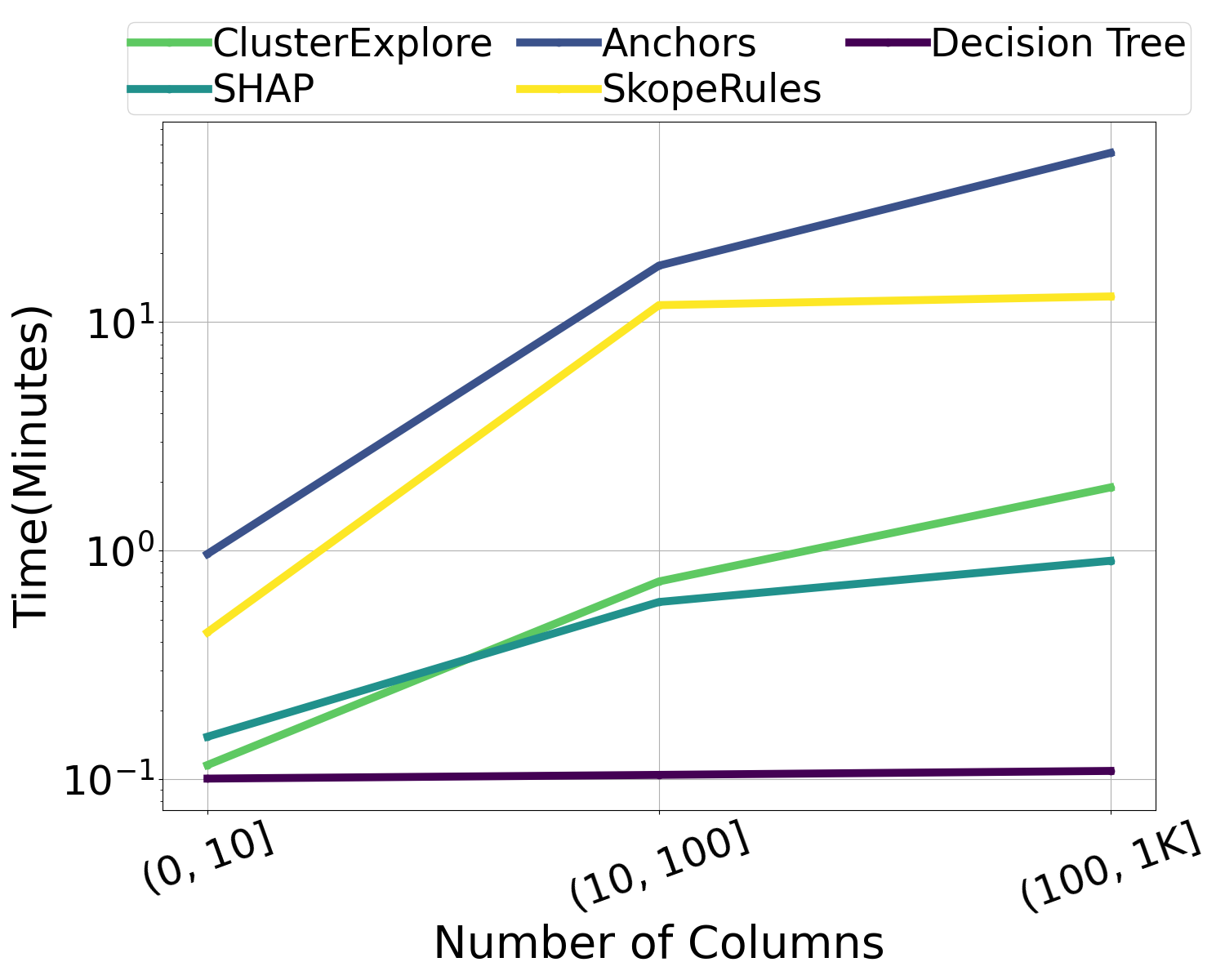}
    \caption{Runtime vs. Number of Columns} 
    \label{fig:time_vs_columns}
    \end{subfigure}
    \begin{subfigure}[b]{.32\linewidth}
    \centering
     \includegraphics[width=\linewidth]{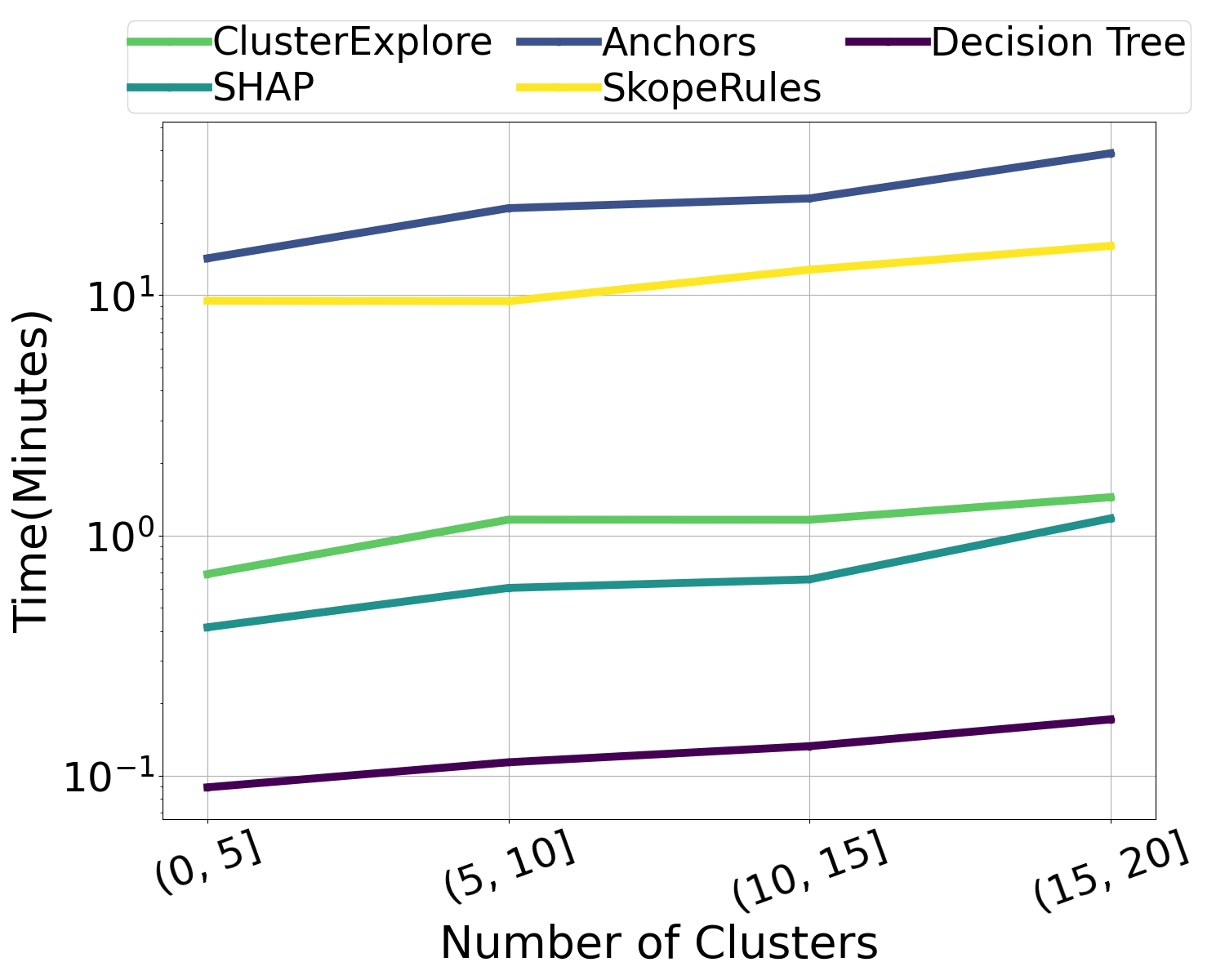}
    \caption{Runtime vs. Number of Clusters}
    \label{fig:time_vs_clusters}
    \end{subfigure}
    \caption{Explanations generation times (for all clusters) as a factor of the number of rows, columns, and clusters.}
    \label{fig:runtime_lines}
\end{figure*}


\subsubsection{Running Times Evaluation}
\label{ssec:runtime}
We further compare the running times of \system{} against the four additional baselines. Figure~\ref{fig:runtime_lines} depicts the average time (in minutes) took for each baseline to return all cluster explanations (i.e., $EX_{ALL}= \{\mathcal{E}_c~|~\forall c \in C \}$), as a factor of the number of rows, columns and resulted clusters. 

Naturally, we observe a running time increase for all baselines as the number of rows increases (Figure~\ref{fig:time_vs_rows}). 
Similar trends are observed for increasing the umber of columns (Figure~\ref{fig:time_vs_columns}), and the number of clusters (Figure~\ref{fig:time_vs_clusters}). 
As expect, the most simple decision tree is indeed the fastest in generating explanations, with an average running time of 6.3 seconds (however, recall that the quality of explanations is significantly subpar compared to SkopeRules and \system{}). 
Also see that the more complex frameworks of SkopeRules and Anchors demonstrate the slowest running times: SkopeRules runs in more than 10 minutes on most settings (except for datasets with less than 1K row and 10 columns), and Anchors often exceeds 20 minutes.

In contrast, the running times of \system{} are significantly better (roughly on par with SHAP), with an average running time of 55.8 seconds. 
In particular, it took \system{} an average of 2.86 minutes to compute all explanations for for datasets larger than 1M rows (compared to, e.g., 24 minutes by SkopeRules).
This is due to the effectiveness of the attribute selection optimization (See Section~\ref{ssec:sampling}) which dramatically decreases running time compared to the exact calculation of \system{} (see below) -- obtaining 12.6X faster times, on average, than the closest baseline in terms of quality.


\begin{figure*}[t!]
    \centering
    \begin{subfigure}[b]{.32\textwidth}
        \centering
        \includegraphics[width=\linewidth]{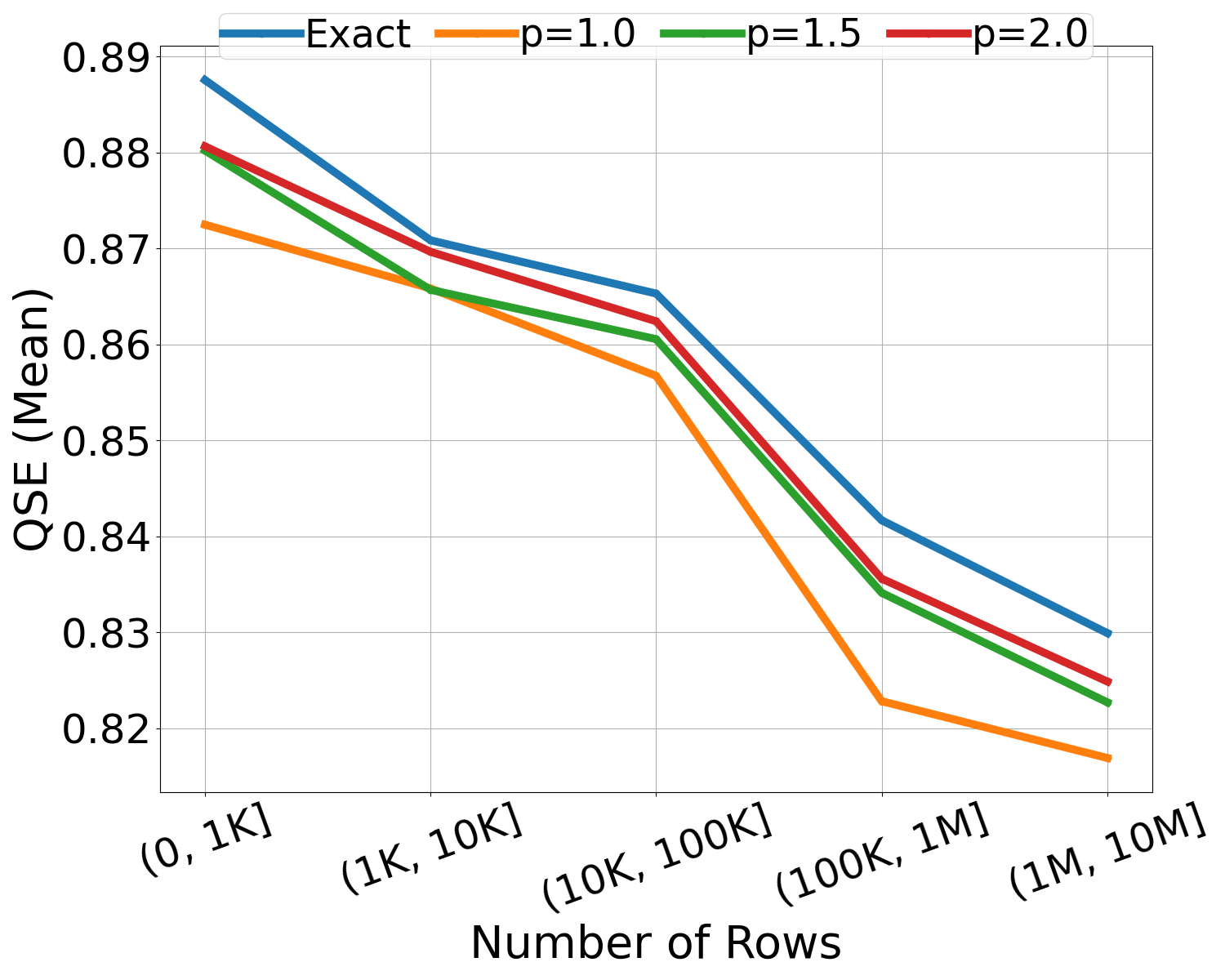}
        \caption{QSE vs. Number of Rows}
        \label{fig:p_sum_mean_vs_rows}
    \end{subfigure}%
    \begin{subfigure}[b]{.32\textwidth}
        \centering
        \includegraphics[width=\linewidth]{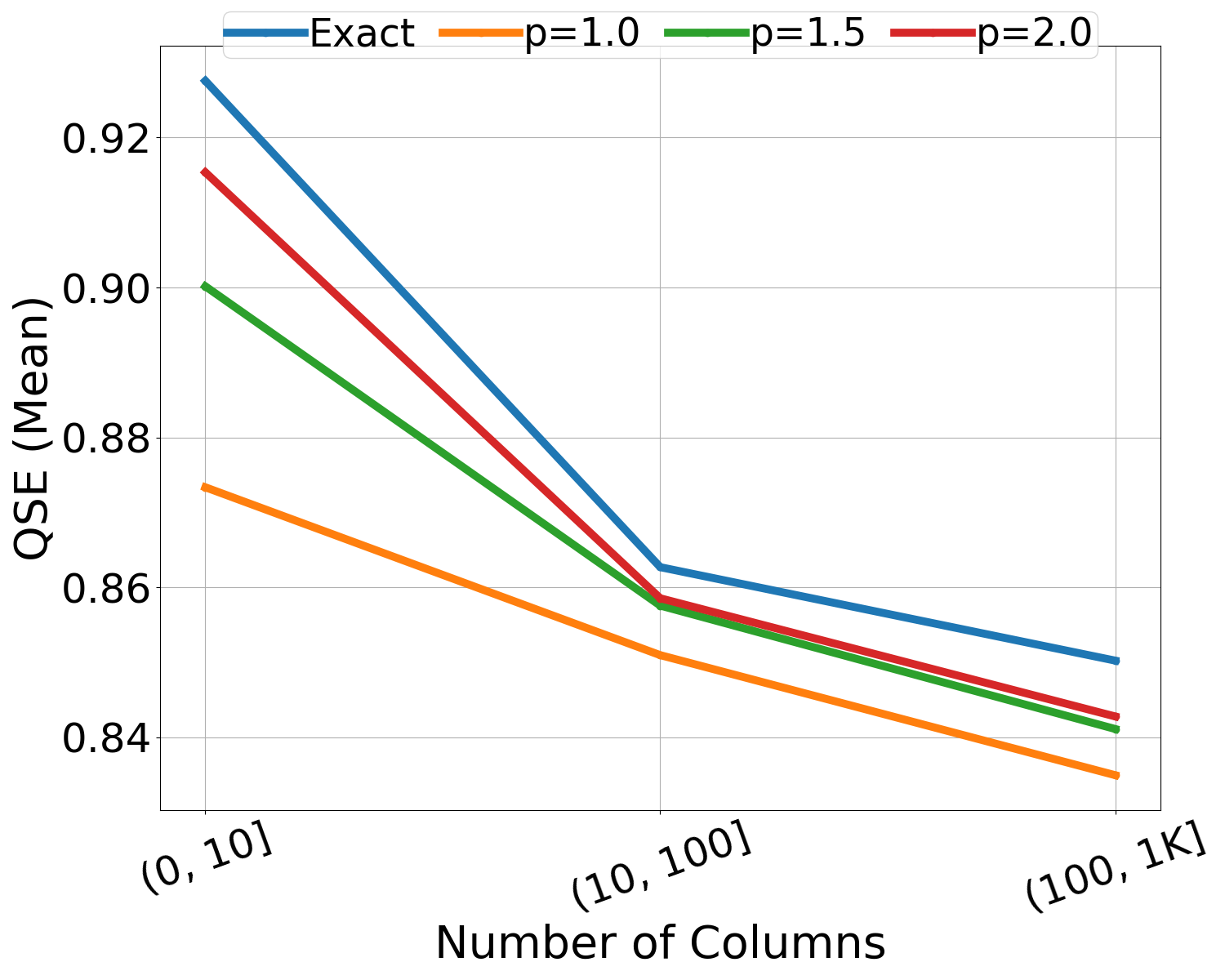}
        \caption{QSE vs. Number of Columns} 
        \label{fig:p_sum_mean_vs_columns}
    \end{subfigure}%
    \begin{subfigure}[b]{.32\textwidth}
        \centering
        \includegraphics[width=\linewidth]{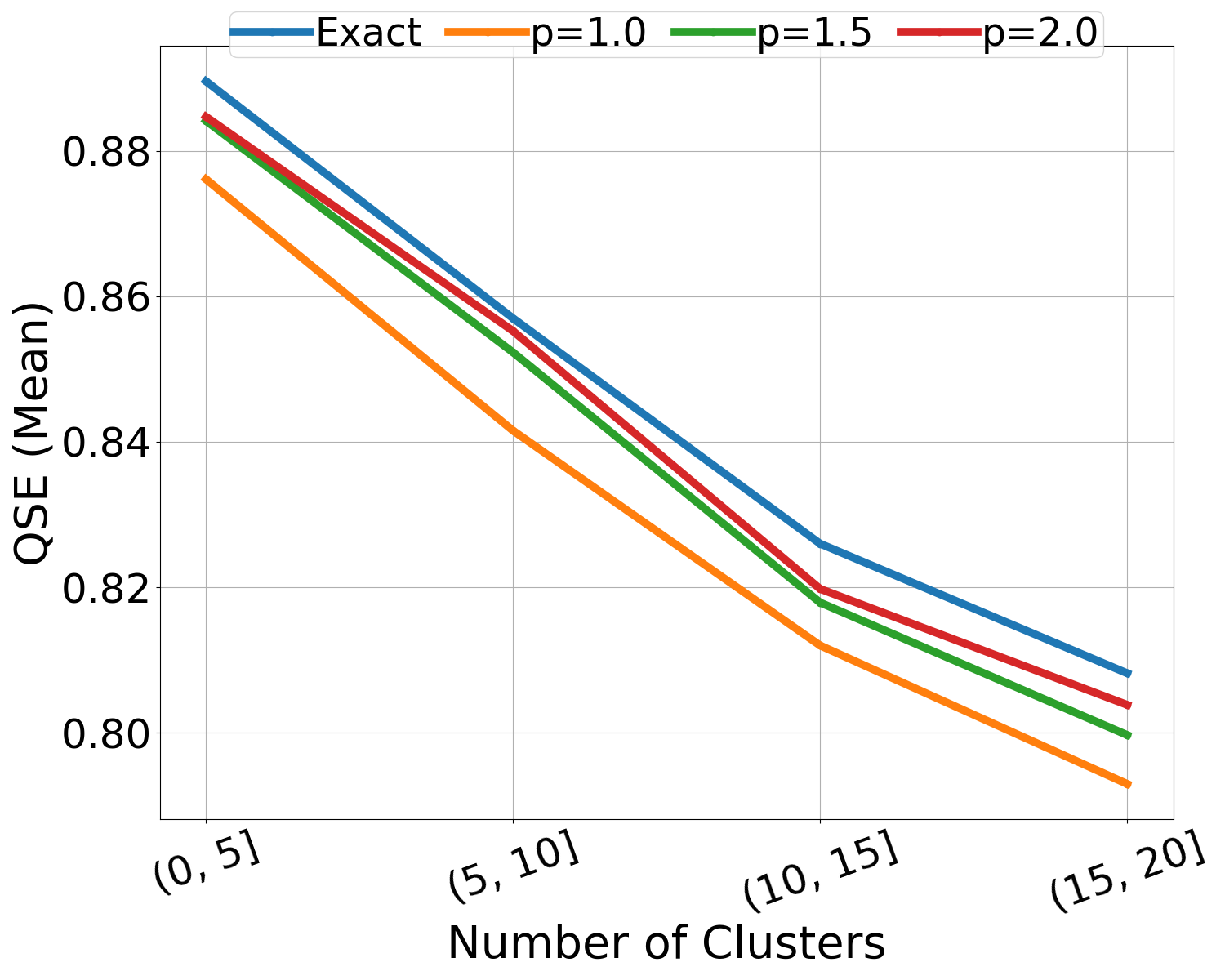}
        \caption{QSE vs. Number of Clusters}
        \label{fig:p_sum_mean_vs_clusters}
    \end{subfigure}
    \caption{The effect of \( p \) on the QSE as a factor of the number of rows, columns, and clusters}
    \label{fig:p_quality_lines}
\end{figure*}

\begin{figure*}[t]
    \centering
    \begin{subfigure}[b]{.32\linewidth}
     \includegraphics[width=\linewidth]{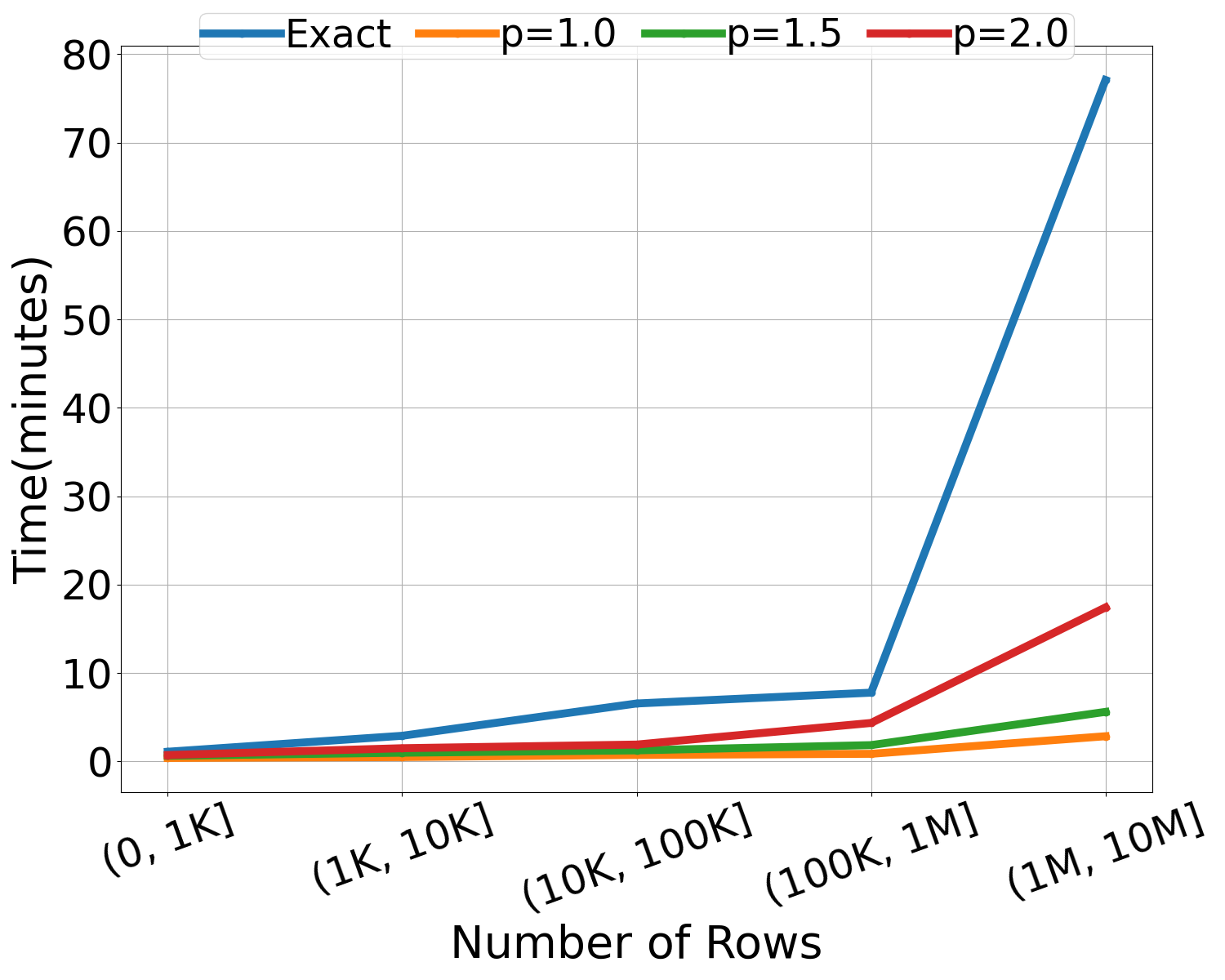}
    \caption{Runtime vs. Number of Rows}
    \label{fig:p_time_vs_rows}
    \end{subfigure}%
    \begin{subfigure}[b]{.32\linewidth}
    \centering
     \includegraphics[width=\linewidth]{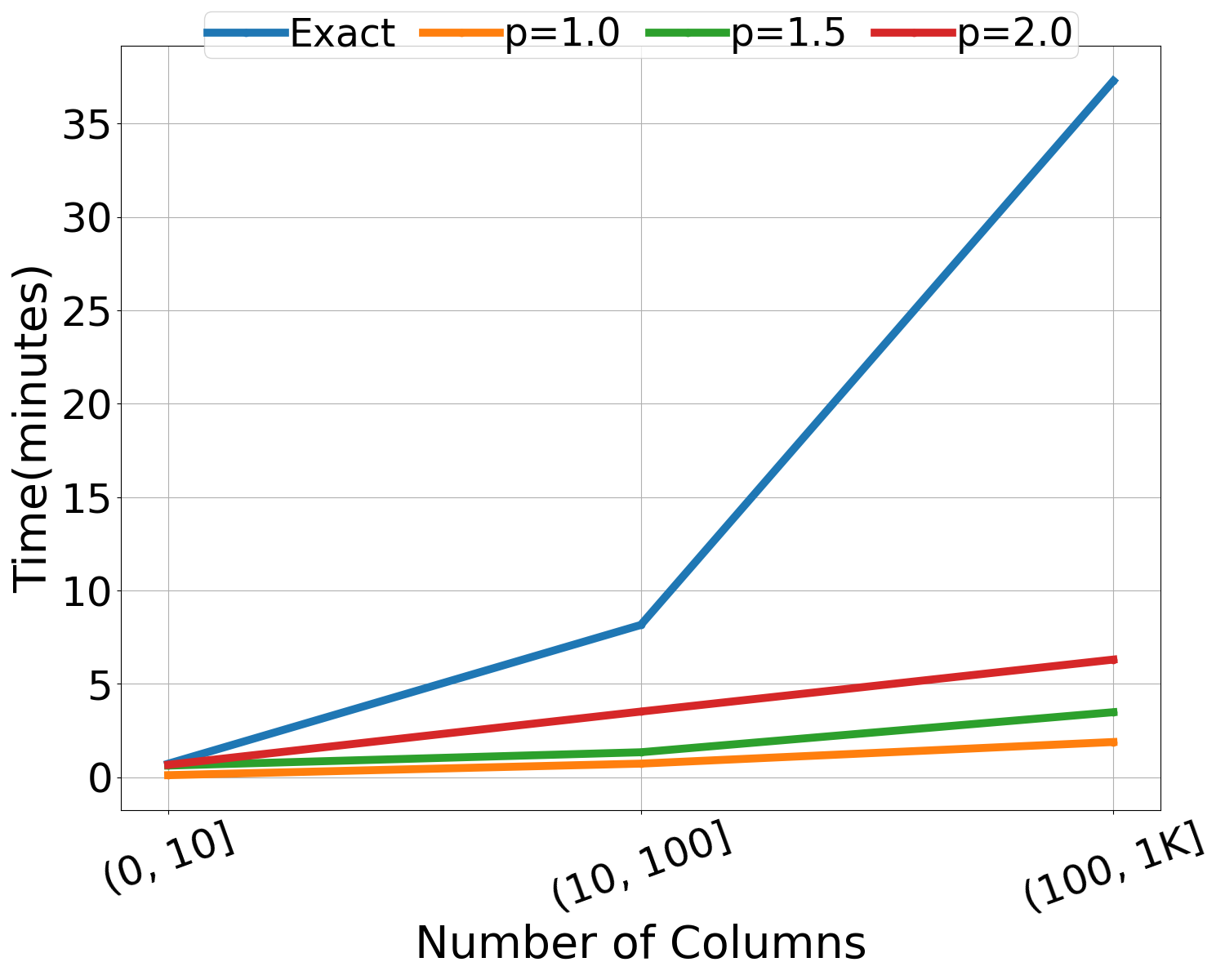}
    \caption{Runtime vs. Number of Columns} 
    \label{fig:p_time_vs_columns}
    \end{subfigure}
    \begin{subfigure}[b]{.32\linewidth}
    \centering
     \includegraphics[width=\linewidth]{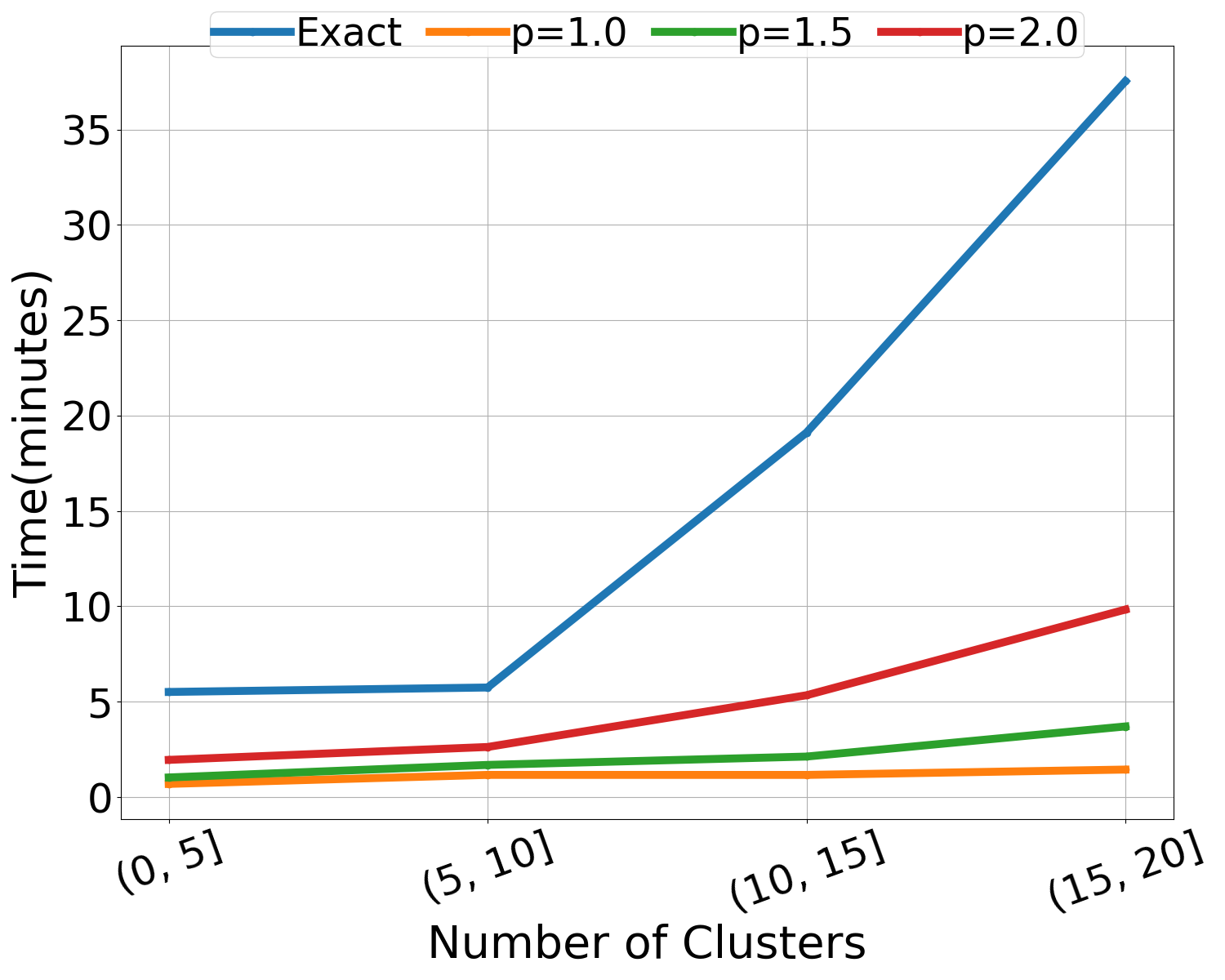}
    \caption{Runtime vs. Number of Clusters}
    \label{fig:p_time_vs_clusters}
    \end{subfigure}
    \caption{The effect of \( p \) on running times as a factor of the number of rows, columns, and clusters}
    \label{fig:p_runtime_lines}
\end{figure*}

\subsubsection{Attribute Selection Effect on Explanations Quality and Generation Times}
\label{ssec:exp_p}

We analyzed the effectiveness of our attribute selection optimization compared to the exact calculation, varying the parameter $p$ (which controls the number of attributes used relative to the conciseness threshold $\theta_{con}$). 

Figure~\ref{fig:p_quality_lines} shows that while the exact \system{} achieves the highest average QSE score (0.86), the optimized versions with $1 \leq p \leq 2$ perform nearly as well, with $p=1$ achieving 0.85 on average. The largest gap occurs for datasets with $\leq 10$ columns, where the exact method scores 0.93 compared to 0.87 for $p=1$, but this difference is less critical due to the shorter runtime of the exact method for small datasets.

In terms of running time, as depicted in Figure~\ref{fig:p_runtime_lines}, the optimization provides substantial speedups for larger datasets. For datasets with over 1M rows, the optimized \system{} achieves a 26.9X improvement, and for datasets with more than 100 columns, the average improvement is 19.7X. Overall, the optimization reduces runtime by 14.4X on average, cutting the exact calculation time from 13.42 minutes to just 55.8 seconds, with minimal impact on QSE scores (<0.1 difference).

%% file: conclusions.tex
\section{conclusion}
\label{sec:conclusion}

\system{} is a novel framework for post-hoc, rule-based explanations of clustering results. It is independent of specific clustering algorithms and avoids using auxiliary ML models for cluster labels. To produce explanations, \system{} first augments the original data with predicates representing numeric intervals and categorical negations, then effectively mines generalized frequent itemsets from the data. An attribute selection optimization further reduces computational cost by limiting items considered. Experiments on 98 clustering results and a user study demonstrate its effectiveness over existing solutions.


In future work, we plan to incorporate disjunctions, \reva{ predict which thresholds to use for coverage and separation error}, \revb{investigate semantic-based evaluation for explanations}, and develop targeted solutions for local explanations as well as for \reva{more complex cluster structures such as hierarchical and density based clustering.} 
